\documentclass[english,aps,nofootinbib,pra,notitlepage]{revtex4-1}
\pdfoutput=1
\usepackage[T1]{fontenc}
\usepackage[latin9]{inputenc}
\usepackage{color}
\usepackage{babel}

\usepackage{amsmath}
\usepackage{graphicx}
\usepackage{subfig}
\usepackage{amssymb}
\usepackage[unicode=true, pdfusetitle,
 bookmarks=false,
 breaklinks=false,pdfborder={0 0 0},backref=false,colorlinks=false]
 {hyperref}

\makeatletter

\providecommand{\tabularnewline}{\\}

\@ifundefined{textcolor}{}
{%
 \definecolor{BLACK}{gray}{0}
 \definecolor{WHITE}{gray}{1}
 \definecolor{RED}{rgb}{1,0,0}
 \definecolor{GREEN}{rgb}{0,1,0}
 \definecolor{BLUE}{rgb}{0,0,1}
 \definecolor{CYAN}{cmyk}{1,0,0,0}
 \definecolor{MAGENTA}{cmyk}{0,1,0,0}
 \definecolor{YELLOW}{cmyk}{0,0,1,0}
 }

\makeatother

\begin{document}

\preprint{Working draft: Not for distribution}

\title{Protecting and Dynamically Generating Entanglement in a Two-Atom Two-Field-Mode Model}

\author{K. Sinha}

\email{kanu@umd.edu}

\author{N. I. Cummings}

\email{nickc@umd.edu}

\author{B. L. Hu}

\email{blhu@umd.edu}

\affiliation{Joint Quantum Institute and Department of Physics\\
University of Maryland, College Park, Maryland 20742-4111, USA}

\date{October 6, 2010}

\begin{abstract}
We analyze the time evolution of quantum entanglement in a model
consisting of two two-level atoms interacting with a two-mode
electromagnetic field for a variety of initial states. We study two
different coupling schemes motivated by the forms that can arise due
to atomic separation. We observe a variety of qualitative features
such as entanglement sudden death, dynamical generation, protection, and transfer between subsystems. Our quantitative analysis shows that
these cases with different couplings and initial states differ
significantly in these qualitative features. The multifarious
behaviors in these two-mode cases suggest the importance of
considering atomic separation carefully for any model where two atoms
interact with a common field.
\end{abstract}

\maketitle
\global\long\def\ket#1{\left|#1\right\rangle }
\global\long\def\bra#1{\left\langle #1\right|}
\global\long\def\tr{\mathrm{Tr}}
\global\long\def\op#1{\hat{#1}}
\global\long\def\defeq{\equiv}
\global\long\def\vect#1{\boldsymbol{#1}}
\global\long\def\leftexp#1#2{{\vphantom{#2}}^{#1}{#2}}

%%%%%%%%%%%%%%%%%%%%%%%%%%%%%%%%%%%%%%%%%%%%%%%%%%%%%%%%%%%%%%%%%%%%%%%%%

\section{Introduction}
Quantum entanglement has been extensively studied, both due to its
fundamental significance in quantum theory \cite{BellIneq,EPR} and
its utility as a resource for quantum communication and quantum
information processing
\cite{NielsenChuangBook,BennettDiVincenzo2000}. Atomic physics offers
a domain with sufficient control of the system and isolation from
noise that it has been possible to do extensive experimental study of
quantum entanglement \cite{OBrien2010,RaimondEtAl2001}, and,
therefore, it is also a productive target for theoretical study of
the issue as well.

One of the simplest scenarios for theoretically studying the dynamics of entanglement
between atoms is that of two atoms which are isolated from one another
and interact with different electromagnetic fields. Studying this
sort of model led to the discovery of the entanglement sudden death
(SD) phenomenon \cite{YuEberly2004,YuEberly2009,Almeida2007,YonacYuEberly2006,YonacYuEberly2007},
in which entanglement decays to zero in a finite time rather than
asymptotically. SD has garnered significant interest because it is
both unintuitive and potentially undesirable.

An alternative simple model for atom-field interaction in which to
study entanglement dynamics is a Dicke model \cite{Dicke},where one
assumes all atoms are grouped in a sample whose size is small
compared to the resonant wavelength (resulting in identical coupling
to every atom). However, such a simplistic model may miss the variety
of behavior that can result when the atoms are not confined to such a
small sample. Entanglement dynamics have been shown to have a
significant distance dependence when two atoms are interacting with a
common field \cite{TaneichiKobayashi2003}. For atoms weakly
interacting with a continuum of field modes in the Born-Markov
approximation, it has been shown
\cite{FicekTanas2008,TanasFicek2004a,TanasFicek2004b,FicekTanis2006}
that changing the atomic separation of two atoms can affect whether
there is SD and whether there is revival of entanglement, as well as
modify the dynamical generation of entanglement; in short, the
qualitative features are sensitive to the atomic spacing. At short
interatomic distance non-Markovian effects associated with induced
interactions between the atoms due to the quantum field become more
pronounced whose qualitative behavior varies greatly with different
classes of initial states, as detailed in the study of
\cite{ASH2006}. So it is clear that distance dependence gives rise to
a significant variety of behaviors, and some seemingly innocuous approximations 
can qualitatively alter the entanglement dynamics in unintuitive ways
\cite{ASH2009,JingLuFicek2009}.

In view of this, we seek to study entanglement dynamics in a model
that is sufficiently complex to manifest some of this variety of
behavior yet simple enough that the dynamics may be understood in
considerable detail and obtained with fewer approximations. The
simplest model one may pursue along this line would be two two-level
atoms (2LA) coupled to a single mode with couplings that reflect the
distance dependence; however, as we demonstrate in Sec.~\ref{sec:setup}, one needs to include at least two field modes before
any non-trivial distance dependence can arise in the problem. So we will 
study such a model with two field modes here. Out of the class of Hamiltonians that can arise from distance
dependence, we will focus on  two special cases with the aim of
illustrating the variety of different behaviors that can arise.
Specifically, we will show how SD, dynamical entanglement generation,
and other phenomena differ between these two cases. We will also
compare with the behavior of the analogous versions of the two
well-studied types of models mentioned above: For two isolated
atom-field systems, we will compare with the case where each system
is a single 2LA interacting with a single field
mode \cite{YonacYuEberly2006,YonacYuEberly2007,Chan2009}. In the
instance of two atoms interacting with the same field, we will
compare with the properties of two 2LAs interacting with a single
common field mode \cite{KimKnight2003,TessierDeutsch2003}. We will
show that the entanglement dynamics in these models differs
significantly from the model of two atoms interacting with a two-mode field that is our focus.

The remainder of this article is structured as follows: In Sec.~\ref{sec:setup} we describe the models we use and the assumptions
behind them. In Sec.~\ref{sec:results} we describe the quantum states of
interest. In Sec.~\ref{sec:EG} we describe the cases where entanglement
is dynamically generated from an initially separable state and
where entanglement is transferred between the atom and the field
sectors. In Sec.~\ref{sec:EP} we describe the cases where
entanglement meets sudden death (SD) and identify the conditions under 
which entanglement can survive.  Situations in which entanglement never dies 
(i.e., the system is never separable) we term always alive (AL). 
In addition to known scenarios of entanglement death,
birth and revival we also encounter situations where entanglement
dies only for an instant (DI). The qualitative features for all the
cases we have studied are summarized in Table \ref{phi0}. Finally in
Sec.~\ref{sec:conclusion} we conclude with a summary and discussion.

%%%%%%%%%%%%%%%%%%%%%%%%%%%%%%%%%%%%%%%%%%%%%%%%%%%%%%%%%%%%%%%%%%%

\section{The Model\label{sec:setup}}

\subsection{System Hamiltonian\label{sec:sysham}}

Consider a pair of two-level atoms (2LA) coupled via the multipolar
interaction Hamiltonian to a collection of electromagnetic (EM) field modes
in the dipole and rotating-wave approximation \cite{AgarwalBook1974}.
The interaction Hamiltonian of the system in the Schr\"{o}dinger
picture has the form \begin{equation}
\op H_{I}=\hbar\sum_{j}\sum_{q}g_{j,q}\op{\sigma}_{j}^{+}\op a_{q}+g_{j,q}^{*}\op{\sigma}_{j}^{-}\op a_{q}^{\dagger},
\end{equation}
 where $\op{\sigma}_{j}^{+}$ and $\op{\sigma}_{j}^{-}$ are the raising and lowering operators for 
the two-level system representing the $j^{\textrm{th}}$ atom, and $\op a_{q}^{\dagger}$ and $\op 
a_{q}$ are the creation and annihilation operators corresponding to the $q^{\textrm{th}}$ normal 
mode with frequency $\omega_{q}$ in some mode decomposition of the EM field with classical 
electric field mode function $\vect f_{q}\left(\vect r\right)$. If we assume the atoms to have 
fixed center of mass positions $\vect R_{j}$,
 assume that the atomic transition for each atom does not alter the angular momentum of the atom, and assume the dipoles associated with all the atomic transitions have identical direction $\vect{e}_{d}$, then  \begin{equation}
g_{j,q}\defeq d_{j}\vect e_{d}\cdot\vect f_{q}\left(\vect R_{j}\right)\sqrt{\frac{\hbar\omega_{q}}{2\epsilon_{0}V}},\label{eq:CouplingForq}\end{equation}
where the field is quantized in a region of volume V and $d_{j}$ is the complex dipole matrix element for the transition in the $j^{\textrm{th}}$ atom \cite{DutraBook2004}.

In general, the mode functions $\vect{f}_{q}\left(\vect R_{j}\right)$ can be quite complicated, 
and even for a single atom there can be position dependence in the dynamics arising from the 
boundary conditions that the mode functions obey. In order to distinguish these effects from the 
effect of atomic separation, we will consider traveling-wave mode functions that are solutions to 
an EM boundary value problem which is translation invariant in the coordinate separating the two 
atoms. For simplicity, we will discuss traveling plane waves in free space, though another 
example would be the modes of a toroidal resonator.

Having selected traveling plane waves as our mode decomposition, we can then choose to define 
them such that for each wave vector $\vect{k}$ one of the polarizations is perpendicular to 
$\vect{e}_d$, so that it does not couple to the problem and can be neglected.  With this 
assumption we know that for all remaining modes \begin{equation} \vect{e}_d \cdot 
\vect{f}_{q}\left(\vect R_{j}\right)=\left\Vert \vect{e}_d \times \frac{\vect{k}}{\left\Vert 
\vect{k} \right\Vert} \right\Vert e^{i\vect k_{q}\cdot\vect R_{j}}.\end{equation} In general each 
$d_{j}$ may have a different complex phase, so \begin{equation} \arg 
\left(g_{j,q}\right)=\arg\left(d_{j}\right)+\vect k_{q}\cdot\vect R_{j}. \end{equation}
 However, it turns out that, without loss of generality, one may study
a much smaller set of Hamiltonians. 

\subsection{Equivalent Hamiltonians\label{sec:equivham}}

Given a particular set of complex phases for the coupling constants
$g_{j,q}$, one may make a trivial basis transformation \begin{equation}
\op U_{b}=\prod_{l}e^{-i\xi_{l}\op{\sigma}_{l}^{z}/2}\prod_{s}e^{i\zeta_{s}\op a_{s}^{\dagger}\op a_{s}},\end{equation}
 which simply amounts to redefining the reference for the phases of
the atoms by $\op{\sigma}_{j}^{+}\rightarrow\op{\sigma}_{j}^{+}e^{-i\xi_{j}}$
and for the fields by $\op a_{q}\rightarrow\op a_{q}e^{-i\zeta_{q}}$.
After this transformation, the interaction Hamiltonian becomes \begin{equation}
\op H_{I}^{\prime}=\op U_{b}\op H_{I}\op U_{b}^{\dagger}=\op H_{I}=\hbar\sum_{j}\sum_{q}g_{j,q}^{\prime}\op{\sigma}_{j}^{+}\op a_{q}+\left(g_{j,q}^{\prime}\right)^{*}\op{\sigma}_{j}^{-}\op a_{q}^{\dagger}\end{equation}
 with \begin{equation}
g_{j,q}^{\prime}=g_{j,q}e^{-i\xi_{j}}e^{-i\zeta_{q}},\end{equation}
and we may obtain the solution to the original problem by using
the transformed Hamiltonian: 
\begin{equation}
\ket{\psi\left(t\right)}=e^{-i\op H_{I}t/\hbar}\ket{\psi\left(0\right)}=\op U_{b}^{\dagger}e^{-i\op H_{I}^{\prime}t/\hbar}\op U_{b}\ket{\psi\left(0\right)}\defeq\op U_{b}^{\dagger}e^{-i\op H_{I}^{\prime}t/\hbar}\ket{\psi^{\prime}\left(0\right)}.\end{equation}
 We may always choose a set of phase shifts $\xi_{l}$ and $\zeta_{s}$
such that for some particular $j$ and $m$ \begin{equation}
\arg\left(g_{l,s}^{\prime}\right)=\arg\left(d_{l}\right)+\vect k_{s}\cdot\vect R_{l}-\xi_{l}-\zeta_{s}=0\quad\textrm{if }l=j\mbox{ or }s=m\end{equation}
 making those couplings real, but for other $l$ and $s$ the phase
difference \begin{equation}
\arg\left(g_{l,s}\right)-\arg\left(g_{j,q}\right)=\arg\left(d_l\right)-\arg\left(d_j\right) + \vect k_{s}\cdot\vect R_{l}-\vect k_{q}\cdot\vect R_{j}\end{equation}
 cannot generally be eliminated by the added phase shifts and
represents the origin of non-trivial physical distance dependence.

If we consider two identical atoms, with frequencies $\omega_{0}$ and dipole strengths 
$\left|d_{j}\right| = d$ coupled to a single EM field mode, then the previous paragraph implies 
that it suffices to consider only a Hamiltonian where both couplings are real, and the atomic 
separation does not enter; thus, there can be no non-trivial distance dependence. However, if we 
consider the two identical atoms coupled to two field modes, then without loss of generality we 
can write the total Hamiltonian of the system as \begin{multline} 
\hat{H}=\hbar\omega_{0}\left(\op{\sigma}_{1}^{\dagger}\op{\sigma}_{1}+\op{\sigma}_{2}^{\dagger}\op{\sigma}_{2}\right)+\hbar\omega_{1}\op 
a_{1}^{\dagger}\op a_{1}+\hbar\omega_{2}\op a_{2}^{\dagger}\op a_{2}+\hbar 
g_{1}\left(\op{\sigma}_{1}^{+}\op a_{1}+\op{\sigma}_{1}^{-}\op a_{1}^{\dagger}\right)\\ +\hbar 
g_{2}\left(\op{\sigma}_{1}^{+}\op a_{2}+\op{\sigma}_{1}^{-}\op a_{2}^{\dagger}\right)+\hbar 
g_{1}\left(\op{\sigma}_{2}^{+}\op a_{1}+\op{\sigma}_{2}^{-}\op a_{1}^{\dagger}\right)+\hbar g_{2} 
e^{i\phi} \left(\op{\sigma}_{2}^{+}\op a_{2}+\op{\sigma}_{2}^{-}\op 
a_{2}^{\dagger}\right),\label{eq:TwoModeHaq}\end{multline}
 where all distance dependence arises from $\phi=\left(\vect k_{2}-\vect k_{1}\right)\cdot\left(\vect R_{2}-\vect R_{1}\right)$.
For simplicity, we will
further assume that $\omega_{1}=\omega_{2}=\omega_{0}$ and $\left\vert\vect{k}_{1}\cdot\vect{e}_d\right\vert=\left\vert\vect{k}_{2}\cdot\vect{e}_d\right\vert$, which by
Eq.~\eqref{eq:CouplingForq} implies $g_{1}=g_{2}\defeq g$. Our aim
is to get a sense of the variety of different entanglement dynamics
that can result from different separations and initial conditions. To that end, we will consider
two special cases, which are arguably extreme cases of the general
model, $\phi=0$ and $\phi=\pi$. 

\subsection{Mapping Equivalent Models and Time Evolution\label{sec:equivmods}}

Since we have assumed the two field modes have the same frequency,
rather than using the original modes $F_{1}$ and $F_{2}$ one could
equally well choose a different mode decomposition of the field where
the two modes are replaced by linear combinations $TF_{1}$ and $TF_{2}$
with annihilation operators \begin{align}
\op A_{1} & =\frac{1}{\sqrt{2}}\left(\op a_{1}+\hat{a}_{2}\right)\label{eq:TransMode1}\\
\op A_{2} & =\frac{1}{\sqrt{2}}\left(\op a_{1}-\hat{a}_{2}\right).\label{eq:TransMode2}\end{align}
In the case of two modes with symmetrical coupling (TMSC), where $\phi=0$,
the interaction Hamiltonian can then be written \begin{equation}
\op H_{I}=\sqrt{2}\hbar g\left[\op{\sigma}_{1}^{+}\op A_{1}+\op{\sigma}_{1}^{-}\op A_{1}^{\dagger}+\op{\sigma}_{2}^{+}\op A_{1}+\op{\sigma}_{2}^{-}\op A_{1}^{\dagger}\right]\label{eq:TMSCTransHaq}\end{equation}
 in terms of these transformed modes, so that the atoms only couple
to $\hat{A}_{1}$ and not $\hat{A}_{2}$. This shows that the Hamiltonian
is equivalent to the model of a single mode symmetrically coupled
(SMSC) to two atoms. If we consider the evolution of the reduced density
matrix of the atom $\op{\rho}_{A}\left(t\right)$ in the TMSC model
it should be the same as in the SMSC model with a properly transformed
initial state. Namely, if the total system (atoms and modes) is in
an initial state described by the density matrix $\op{\chi}_{TMSC}\left(0\right)$,
then the appropriate initial density matrix for the SMSC model is obtained
by making the mode transformation of Eqs. \eqref{eq:TransMode1} and
\eqref{eq:TransMode2} and tracing out the second transformed mode
$TF_{2}$. For example, if the initial state in the TMSC model is separable
with the field in a product of Glauber coherent states $\ket{\alpha,\beta}$,
then \begin{multline}
\op{\chi}_{TMSC}\left(0\right)=\ket{\phi}\bra{\phi}\otimes\ket{\alpha,\beta}\bra{\alpha,\beta}\rightarrow\tr_{TF_{2}}\left[\ket{\phi}\bra{\phi}\otimes\ket{\frac{\alpha+\beta}{\sqrt{2}},\frac{\alpha-\beta}{\sqrt{2}}}\bra{\frac{\alpha+\beta}{\sqrt{2}},\frac{\alpha-\beta}{\sqrt{2}}}\right]\\
=\ket{\phi}\bra{\phi}\otimes\ket{\frac{\alpha+\beta}{\sqrt{2}}}\bra{\frac{\alpha+\beta}{\sqrt{2}}}=\op{\chi}_{SMSC}\left(0\right)\end{multline}
 is the appropriate mapping to the equivalent SMSC problem. It is
important to note that, because this mapping of initial states involves
a partial trace, it is a many-to-one mapping from the TMSC problem
to the SMSC problem (and this mapping does not preserve purity). In order to
solve the dynamics in the SMSC model, and by extension the TMSC model,
we simply compute the time evolution operator expressed in the atomic basis $\{\ket{ee},\ket{eg},\ket{ge},\ket{gg}\}$ directly by exponentiation
(as in, e.g. \cite{KimKnight2003}):\[
\hat{H}_{I}=\sqrt{2}g\left(\begin{array}{cccc}
0 & \hat{a} & \hat{A}_{1} & 0\\
\hat{A}^{\dagger}_1 & 0 & 0 & \hat{A}_1\\
\hat{A}^{\dagger}_1 & 0 & 0 & \hat{A}_1\\
0 & \hat{A}^{\dagger}_1 & \hat{A}^{\dagger}_1 & 0\end{array}\right)\Rightarrow\hat{U}=e^{-i\hat{H}_{I}t}=\left(\begin{array}{cccc}
\hat{C}_{1} & -i\hat{S}_{1} & -i\hat{S}_{1} & \hat{C}_{2}\\
-i\hat{S}_{2} & \hat{C}_{3} & \hat{C}_{4} & -i\hat{S}_{3}\\
-i\hat{S}_{2} & \hat{C}_{4} & \hat{C}_{3} & -i\hat{S}_{3}\\
\hat{C}_{5} & -i\hat{S}_{4} & -i\hat{S}_{4} & \hat{C}_{6}\end{array}\right)\]
 where 
\begin{align}
\hat{S}_{1} & =\hat{A}_{1}\frac{\sin\left(\sqrt{4\hat{\mathcal{A}}}gt\right)}{\sqrt{2\hat{\mathcal{A}}}} &
\hat{C}_{1} & =1-\hat{A}_{1}\frac{1}{\hat{\mathcal{A}}}\hat{A}_{1}^{\dagger}+\hat{A}_{1}\frac{\cos\left(\sqrt{4\hat{\mathcal{A}}}gt\right)}{\hat{\mathcal{A}}}\hat{A}_{1}^{\dagger} &
\hat{C}_{5} & =-\hat{A}_{1}^{\dagger}\frac{1}{\hat{\mathcal{A}}}\hat{A}_{1}^{\dagger}+\hat{A}_{1}^{\dagger}\frac{\cos\left(\sqrt{4\hat{\mathcal{A}}}gt\right)}{\hat{\mathcal{A}}}\hat{A}_{1}^{\dagger} \nonumber \\
\hat{S}_{2} & =\frac{\sin\left(\sqrt{4\hat{\mathcal{A}}}gt\right)}{\sqrt{2\hat{\mathcal{A}}}}\hat{A}_{1}^{\dagger} &
\hat{C}_{2} & =-\hat{A}_{1}\frac{1}{\hat{\mathcal{A}}}\hat{A}_{1}+\hat{A}_{1}\frac{\cos\left(\sqrt{4\hat{\mathcal{A}}}gt\right)}{\hat{\mathcal{A}}}\hat{A}_{1} &
\hat{C}_{6} & =1-\hat{A}_{1}^{\dagger}\frac{1}{\hat{\mathcal{A}}}\hat{A}_{1}+\hat{A}_{1}^{\dagger}\frac{\cos\left(\sqrt{4\hat{\mathcal{A}}}gt\right)}{\hat{\mathcal{A}}}\hat{A}_{1} \nonumber \\
\hat{S}_{3} & =\frac{\sin\left(\sqrt{4\hat{\mathcal{A}}}gt\right)}{\sqrt{2\hat{\mathcal{A}}}}\hat{A}_{1} &
\hat{C}_{3} & =\frac{1}{2}\left(\cos\left(\sqrt{4\hat{\mathcal{A}}}gt\right)+1\right) &
\hat{\mathcal{A}} & \equiv\hat{A}_{1}\hat{A}_{1}^{\dagger}+\hat{A}_{1}^{\dagger}\hat{A}_{1} \\
\hat{S}_{4} & =\hat{A}_{1}^{\dagger}\frac{\sin\left(\sqrt{4\hat{\mathcal{A}}}gt\right)}{\sqrt{2\hat{\mathcal{A}}}} &
\hat{C}_{4} & =\frac{1}{2}\left(\cos\left(\sqrt{4\hat{\mathcal{A}}}gt\right)-1\right). & &\nonumber 
\end{align}

When $\phi=\pi$ we have a two-mode model with asymmetrical coupling
(TMAC), and we may again use the mode transformation of Eqs.~\eqref{eq:TransMode1}
and \eqref{eq:TransMode2} and write the Hamiltonian as \begin{equation}
\op H_{I}=\sqrt{2}\hbar g\left[\op{\sigma}_{1}^{+}\op A_{1}+\op{\sigma}_{1}^{-}\op A_{1}^{\dagger}+\op{\sigma}_{2}^{+}\op A_{2}+\op{\sigma}_{2}^{-}\op A_{2}^{\dagger}\right].\label{eq:TMACTransHaq}\end{equation}
 In this case, rather than both atoms coupling to one mode we see
that atom one couples only to transformed mode $TF_{1}$ while atom
two couples only to $TF_{2}$, implying that this Hamiltonian is equivalent
to a model comprised of two subsystems that are totally isolated from
one another, each composed of a single atom coupled to a single mode.
We will call this the double Jaynes-Cummings (DJC) model. This sort
of model with isolated subsystems is common to the study of entanglement
sudden death \cite{YuEberly2004,YuEberly2009}, and the DJC model
specifically has been studied \cite{Chan2009,YonacYuEberly2006,YonacYuEberly2007}.
As in the previous case, the evolution of $\op{\rho}_{A}\left(t\right)$
for the TMAC model should be the same as given by the DJC model with
the proper mapping of initial states. In this case the mapping of
initial states is limited to transforming the modes according to
Eqs.~\eqref{eq:TransMode1} and \eqref{eq:TransMode2}, so \begin{equation}
\ket{\phi}\otimes\ket{\alpha,\beta}\rightarrow\ket{\phi}\otimes\ket{\frac{\alpha+\beta}{\sqrt{2}},\frac{\alpha-\beta}{\sqrt{2}}}.\end{equation}
This mapping implies that there can, for example, be no dynamical
generation of atomic entanglement in the TMAC model unless the DJC initial
field state obtained by the mapping is entangled. In the DJC model we can write the
unitary time-evolution operators for the two separate non-interacting atom-field
subsystems as $\hat{U}_{1}$ and $\hat{U}_{2}$, and then the total
time evolution operator is $\op U=\op U_{1}\otimes\op U_{2}$. We
again compute the two subsystem unitary time evolution operators by direct exponentiation
to obtain 
\begin{align}
\hat{U}_{1}=e^{-i\hat{H}_{1}t}= & \left(\begin{array}{cccc}
\hat{C}_{11} & 0 & -i\hat{S}_{11} & 0\\
0 & \hat{C}_{11} & 0 & -i\hat{S}_{11}\\
-i\hat{S}_{12} & 0 & \hat{C}_{12} & 0\\
0 & -i\hat{S}_{12} & 0 & \hat{C}_{12}\end{array}\right)\\
\hat{U}_{2}=e^{-i\hat{H}_{2}t}= & \left(\begin{array}{cccc}
\hat{C}_{21} & -i\hat{S}_{21} & 0 & 0\\
-i\hat{S}_{21} & \hat{C}_{21} & 0 & 0\\
0 & 0 & \hat{C}_{22} & -i\hat{S}_{22}\\
0 & 0 & -i\hat{S}_{22} & \hat{C}_{22}\end{array}\right)\end{align}
 with \begin{align}
\hat{C}_{i1} & =\cos(\sqrt{2\hat{A}_{i}\hat{A}_{i}^{\dagger}}gt) & 
\hat{C}_{i2} & =\cos(\sqrt{2\hat{A}_{i}^{\dagger}\hat{A}_{i}}gt) \\
\hat{S}_{i1} & =\frac{\sin(\sqrt{2\hat{A}_{i}\hat{A}_{i}^{\dagger}}gt)}{\sqrt{\hat{A}_{i}\hat{A}_{i}^{\dagger}}}\hat{A}_{i} & 
\hat{S}_{i2} & =\hat{A}_{i}^{\dagger}\frac{\sin(\sqrt{2\hat{A}_{i}^{\dagger}\hat{A}_{i}}gt)}{\sqrt{\hat{A}_{i}^{\dagger}\hat{A}_{i}}}.\nonumber 
\end{align}

\subsection{Quantifying Entanglement}

In order to quantitatively study the entanglement dynamics, we must specify a way of quantifying 
entanglement. For two qubits one may compute the entanglement measure known as the entanglement 
of formation efficiently in terms of the Wootters concurrence \cite{Wooters1998}. In order to 
compute the concurrence, one must choose an arbitrary basis and construct the spin-flipped 
operator \begin{equation} 
\tilde{\rho}\defeq\left(\hat{\sigma}_{y}\otimes\hat{\sigma}_{y}\right)\hat{\rho}^{*}\left(\hat{\sigma}_{y}\otimes\hat{\sigma}_{y}\right),\end{equation}
 where the star denotes the complex conjugation in that basis and
$\hat{\sigma}_{y}$ is the Pauli matrix in the same basis. The concurrence
is \begin{equation}
C\left(\op{\rho}\right)=\max\left\{ 0,\sqrt{\lambda_{1}}-\sqrt{\lambda_{2}}-\sqrt{\lambda_{3}}-\sqrt{\lambda_{4}}\right\} ,\end{equation}
 where the $\lambda_{j}$s are the eigenvalues of the matrix $\op{\rho}\tilde{\rho}$
in decreasing order. The entanglement of formation can then be computed
as \begin{equation}
\mathcal{E}_{F}\left(C\right)=h\left(\frac{1+\sqrt{1-C^{2}}}{2}\right),\end{equation}
 where \begin{equation}
h\left(x\right)=-x\log_{2}\left(x\right)-\left(1-x\right)\log_{2}\left(1-x\right).\end{equation}
 Alternatively, to quantify entanglement one can also use a monotone known as the negativity \cite{VidalWerner2002} (which quantifies the degree to which the state violates the Peres-Horodecki positive partial transpose condition), defined
\begin{equation}
\mathcal{N}\left(\op{\rho}\right)=\frac{\left\Vert \hat{\rho}^{T_{B}}\right\Vert _{1}-1}{2},
\end{equation}
where $\hat{\rho}^{T_{B}}$ is the partial transpose of the density
matrix, defined by \begin{equation}
\bra{j,k}\hat{\rho}^{T_{B}}\ket{l,q}\defeq\bra{j,q}\hat{\rho}\ket{l,k}\end{equation}
 in an arbitrary basis B, and $\left\Vert \cdot\right\Vert _{1}$ is the trace norm. Either of 
these approaches can be used to quantify entanglement.  The negativity is arguably less precise, 
because in some systems it can be zero even for an entangled state, but we will not consider such 
situation.  The advantage of the negativity is that it can be efficiently computed even in a 
large Hilbert space.  We will most often use the concurrence in this article; however, we do 
quantify the entanglement of field states using the negativity.

\section{The States\label{sec:results}}

In order to illustrate the variety of behavior that can arise among
the four models we discussed in Sec.~\ref{sec:equivmods}, we will examine
the entanglement dynamics of a selection of initial states in which
the atoms are separable from the fields and which are comprised of
familiar atomic and field states. In each case we will select the
atomic state from the set of pure states $\left\{ \ket{gg},\ket{ee},\ket{eg},\ket{\Phi},\ket{\Psi}\right\} $,
where $\ket{\Phi}\defeq\left(\ket{ee}+\ket{gg}\right)/\sqrt{2}$,
$\ket{\Psi}\defeq\left(\ket{eg}+\ket{ge}\right)/\sqrt{2}$, and $e$
and $g$ label the excited and ground states of the atom respectively.
The initial field state will be either the vacuum $\ket{00}$, a product
of Fock states $\ket{n_N,m_{N}}$, a product of Glauber coherent
states $\ket{\alpha_{c},\beta_{c}}$, a product of squeezed vacuum
states $\ket{\xi_{sq},-\xi_{sq}}$ , a two-mode squeezed vacuum state
(TMSS) $\ket{\xi,0,0}$ , a thermal state $\op{\rho}_{th}$ (with
both modes having equal temperature), the pure state $\ket{\eta_{nm}}$,
or the mixed state $\op{\rho}_{nm}$. The two-mode squeezed state is defined as the state resulting from the action of the two-mode squeezing operator $\hat{S}(\xi)=e^{\left(\xi^{\ast}\hat{a}_1\hat{a}_2-\xi\hat{a}_1^{\dagger}\hat{a}_2^{\dagger}\right)}$ on vacuum. The state $\ket{\eta_{nm}}$ is
the result of mapping the state $\ket{n_{N},m_{N}}$ in the original
modes of the TMAC problem to the transformed modes equivalent to
the DJC problem, with \begin{equation}
\ket{\eta_{nm}}=\frac{1}{\sqrt{2^{m+n}m!n!}}\sum_{k=0}^{n}{\sum_{l=0}^{m}{\leftexp nC_{k}\leftexp mC_{l}\sqrt{(m+n-k-l)!(k+l)!}(-1)^{l}}}\ket{m+n-k-l}\ket{k+l},\end{equation}
and $\hat{\rho}_{nm}\defeq\tr_{TF_{2}}\left[\ket{\eta_{nm}}\bra{\eta_{nm}}\right]$ is the state in the SMSC model that gives equivalent evolution to the state $\ket{n_{N},m_{N}}$ in the TMSC model. $\leftexp nC_{k}$ represents the binomial coefficients.

The correspondence drawn between the TMSC-SMSC leads to the essential feature that the map for the initial field states from TMSC to SMSC is many to one; the set of initial states that
have the same reduced density matrix for the first transformed
mode $TF_{1}$ have identical entanglement dynamics in terms of $A_{1}$-$A_{2}$
entanglement. As a counterintuitive example of this feature we will see
that a squeezed state of the form $\ket{\xi_{sq},-\xi_{sq},0}$ and
a thermal field in the TMSC model give the same entanglement dynamics provided the
average number of photons in the thermal field corresponds to that
in the squeezed state ($\bar{n}_{th}=\sinh^{2}\left\vert \xi_{sq}\right\vert $). 

We summarize
our findings for the entanglement behavior given the various initial
states considered in the four models in Table \ref{phi0}, listing
the equivalent TMSC-SMSC cases and
 TMAC-DJC cases.
When discussing entanglement sudden death, we adopt the usage of the term
as in \cite{YonacYuEberly2006} in applying it only to instances where
the entanglement goes to zero for some time interval of non-zero length. In the case where entanglement goes to zero only for an instant during the time evolution we refer to it as death for an instant (DI). If there is a non-zero entanglement at all times once it is generated in the system then we label it as being ``always living'' (AL).	

\begin{table}
\begin{tabular}{|l|l|c|c|c|c|c|}
\hline 
\multicolumn{2}{|c}{Initial field State} & \multicolumn{5}{|c|}{Atomic State}\tabularnewline
\hline 
TMSC  & SMSC  & A. $\ket{ee}$  & B. $\ket{eg}$  & C. $\ket{gg}$  & D. $\ket{\Phi}$  & E. $\ket{\Psi}$\tabularnewline
\hline
\hline 
1. $\ket{00}$  & $\ket{0}$  & No  & Yes, DI  & Yes, DI  & AL  & DI \tabularnewline
\hline 
2. $\ket{n_{N},m_{N}}$  & $\hat{\rho}_{nm}$  & No  & Yes, DI  & Yes\footnote[1]{No entanglement for $n_N=m_N$}, DI/SD  & SD  & SD\tabularnewline
\hline 
3. $\ket{\eta_{nm}}$ & $\ket{n_{N}}$  & No  & Yes, DI  & Yes, SD  & SD  & SD\tabularnewline
\hline 
4. $\hat{\rho}_{th}$  & $\hat{\rho}_{th}$  & No  & Yes, DI  & Yes, SD  & AL/SD & SD \tabularnewline
\hline 
5. $\ket{\frac{1}{\sqrt{2}}\left(\alpha_{c}+\beta_{c}\right),\frac{1}{\sqrt{2}}\left(\alpha_{c}-\beta_{c}\right)}$  & $\ket{\alpha_{c}}$  & Yes, AL/SD  & Yes, SD  & Yes, AL/SD  & AL/SD  & AL/SD \tabularnewline
\hline 
6. $\ket{\xi_{sq},-\xi_{sq}}$  & $\hat{\rho}_{th}$  & No  & Yes, DI  & Yes, SD  & AL/SD & SD \tabularnewline
\hline 
7. $\ket{\xi,0,0}$  & $\ket{\xi_{sq}}$  & Yes, AL/SD  & Yes,SD  & Yes, AL/SD  & AL  & SD \tabularnewline
\hline
\multicolumn{1}{c}{} & \multicolumn{1}{c}{} & \multicolumn{1}{c}{} & \multicolumn{1}{c}{} & \multicolumn{1}{c}{} & \multicolumn{1}{c}{} & \multicolumn{1}{c}{}\tabularnewline
\hline 
\multicolumn{2}{|c}{Initial field State} & \multicolumn{5}{|c|}{Atomic State}\tabularnewline
\hline 
TMAC  & DJC  & A. $\ket{ee}$  & B. $\ket{eg}$  & C. $\ket{gg}$  & D. $\ket{\Phi}$  & E. $\ket{\Psi}$\tabularnewline
\hline
\hline 
1. $\ket{00}$  & $\ket{00}$  & No  & No  & No  & SD  &DI\tabularnewline
\hline 
2. $\ket{n_{N},m_{N}}$  & $\ket{\eta_{nm}}$  & Yes\footnotemark[1], SD  & Yes\footnotemark[1], SD  & Yes\footnotemark[1], SD/DI  & SD  & SD/AL \tabularnewline
\hline 
3. $\ket{\eta_{nm}}$  & $\ket{n_{N},m_{N}}$  & No  & No  & No  & SD  & SD \tabularnewline
\hline 
4. $\hat{\rho}_{th}$  & $\hat{\rho}_{th}$  & No  & No  & No  & SD  & SD \tabularnewline
\hline 
5. $\ket{\alpha_{c},\beta_{c}}$  & $\ket{\frac{1}{\sqrt{2}}\left(\alpha_{c}+\beta_{c}\right),\frac{1}{\sqrt{2}}\left(\alpha_{c}-\beta_{c}\right)}$  & No  & No  & No  & SD  & SD\tabularnewline
\hline 
6. $\ket{\xi_{sq},-\xi_{sq}}$  & $\ket{\xi,0,0}$  & Yes, SD  & Yes, SD  & Yes, SD  & SD  & SD \tabularnewline
\hline 
7. $\ket{\xi,0,0}$  & $\ket{\xi_{sq},-\xi_{sq}}$  & No  & No  & No  & SD  & SD \tabularnewline
\hline
\end{tabular}

\caption{Entanglement dynamics for two modes symmetrical coupling  (TMSC) with $\phi=0$ and anti-symmetrical coupling with $\phi=\pi$. Columns A-C list whether there is an entanglement generation in an initially separable atomic state (yes or no). The dynamical phenomena observed in columns A-E are listed as entanglement sudden death (SD), entanglement dies for only an instant (DI), entanglement remains non-zero at all times and so is ``always living'' (AL). A `/' denotes that both kinds of dynamics are present depending on the particular initial state chosen from within the class indicated. The initial states have been explained in Sec.~\ref{sec:results}.}
\label{phi0}
\end{table}

\section{Entanglement Generation and Transfer\label{sec:EG}}

The generation of entanglement from an initially separable state by
the dynamics of the system is one interesting, and potentially useful, phenomenon to examine.  
In the DJC model each atom interacts only with a separate field mode, so the dynamics cannot increase entanglement between the two atom-field subsystems; therefore, if the atomic state is not entangled initially then it will remain separable, unless there is an initial entanglement between the field modes that can be transfered to the atoms by the dynamics.  Knowing this, we can see that any initial field state for the TMAC model that maps to a separable DJC field state will also fail to generate entanglement.  The nature of the mapping means that even some entangled field states will fail to generate atomic entanglement in the TMAC model, while some separable states will map to an entangled DJC state and will generate entanglement.  

\begin{figure}[ht]
\subfloat[$\ket{ee}$ - No SD once entanglement is generated if the state is sufficiently squeezed]{\label{ssqee}\includegraphics[width=3.2 in]{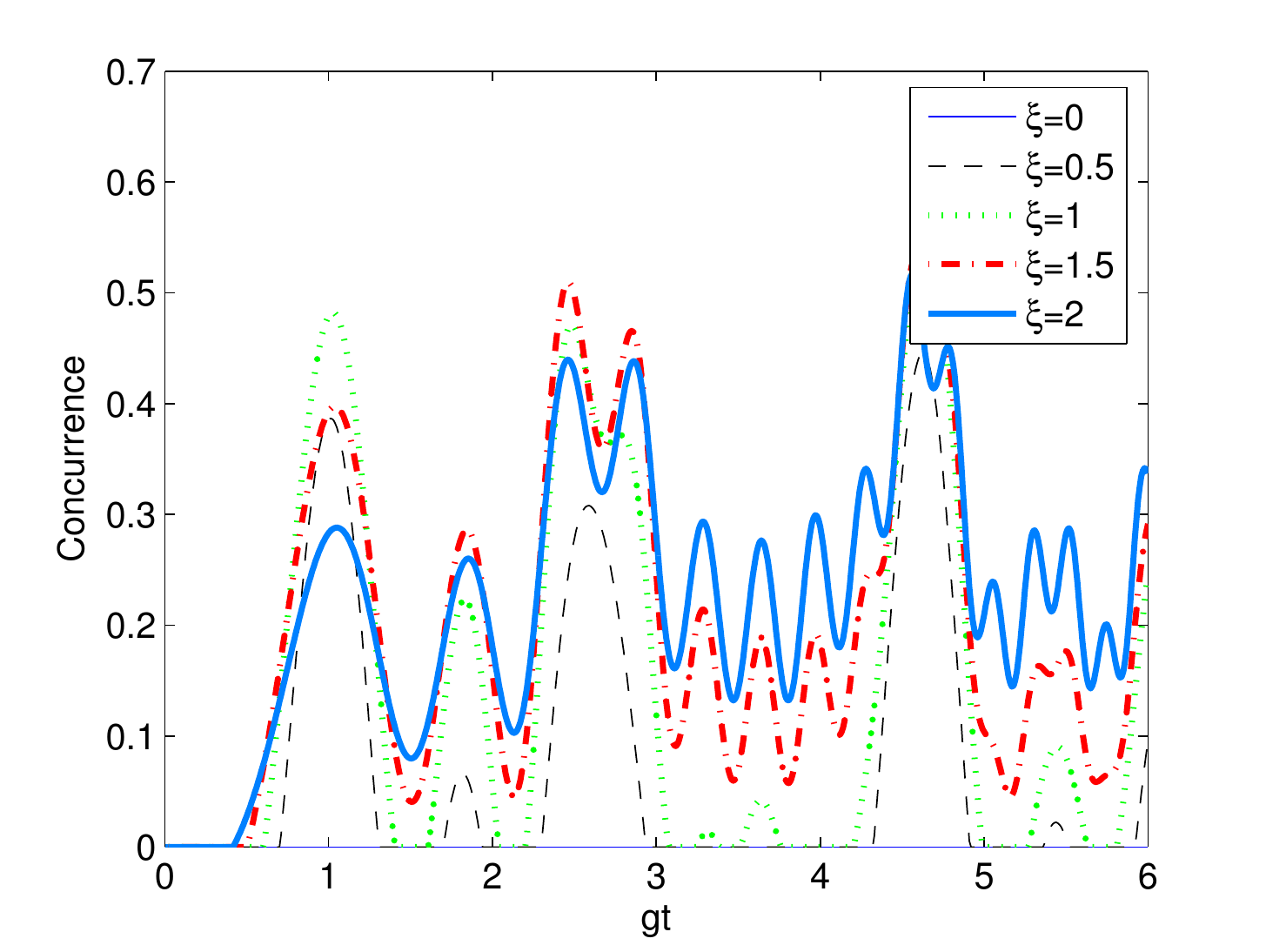}}
\subfloat[$\ket{eg}$ - increased squeezing destroys entanglement in this case]{\label{ssqeg}\includegraphics[width=3.2 in]{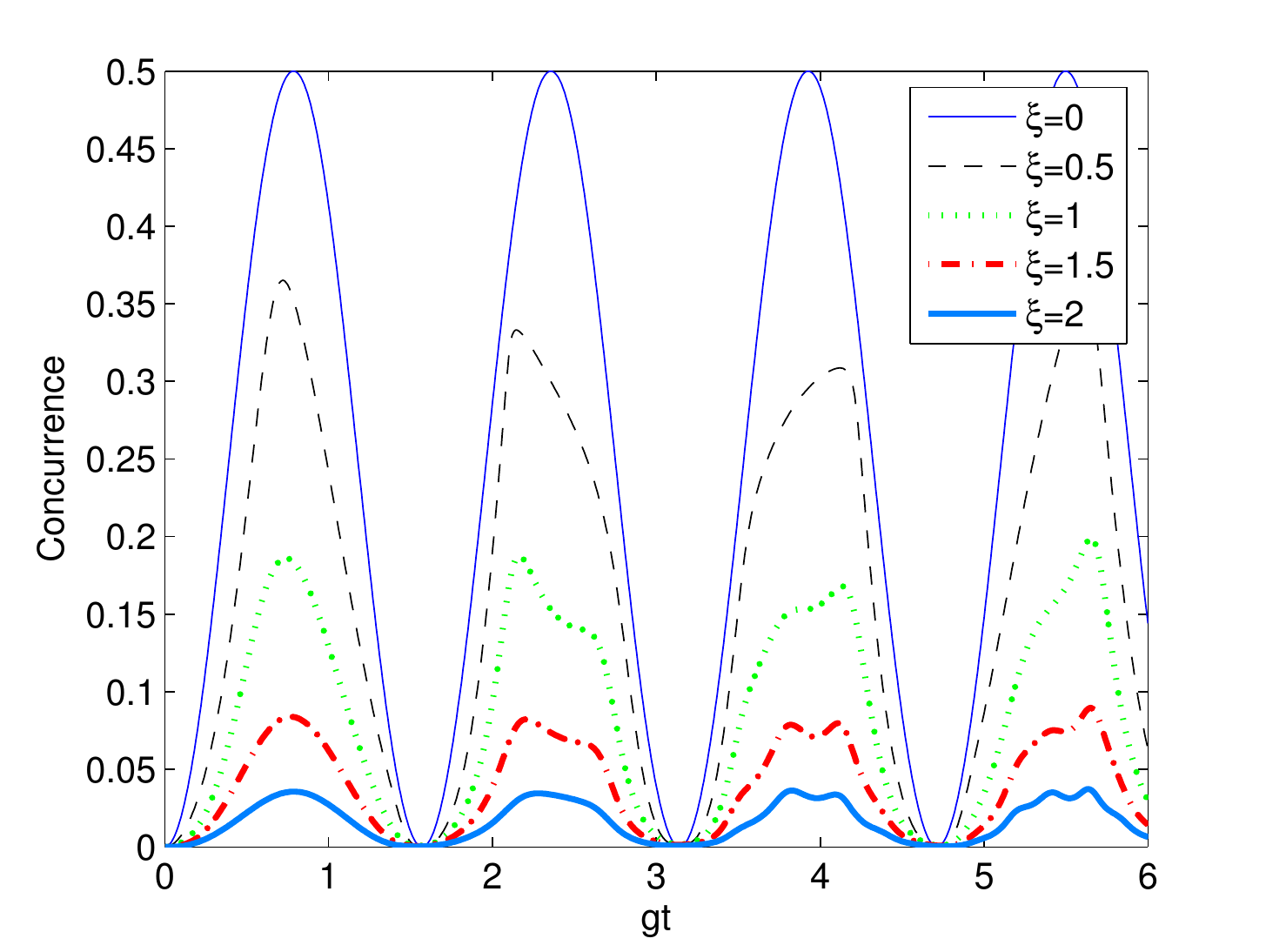}}\\
\subfloat[$\ket{gg}$ - SD disappears on increasing the squeezing]{\label{ssqgg}\includegraphics[width=3.2 in]{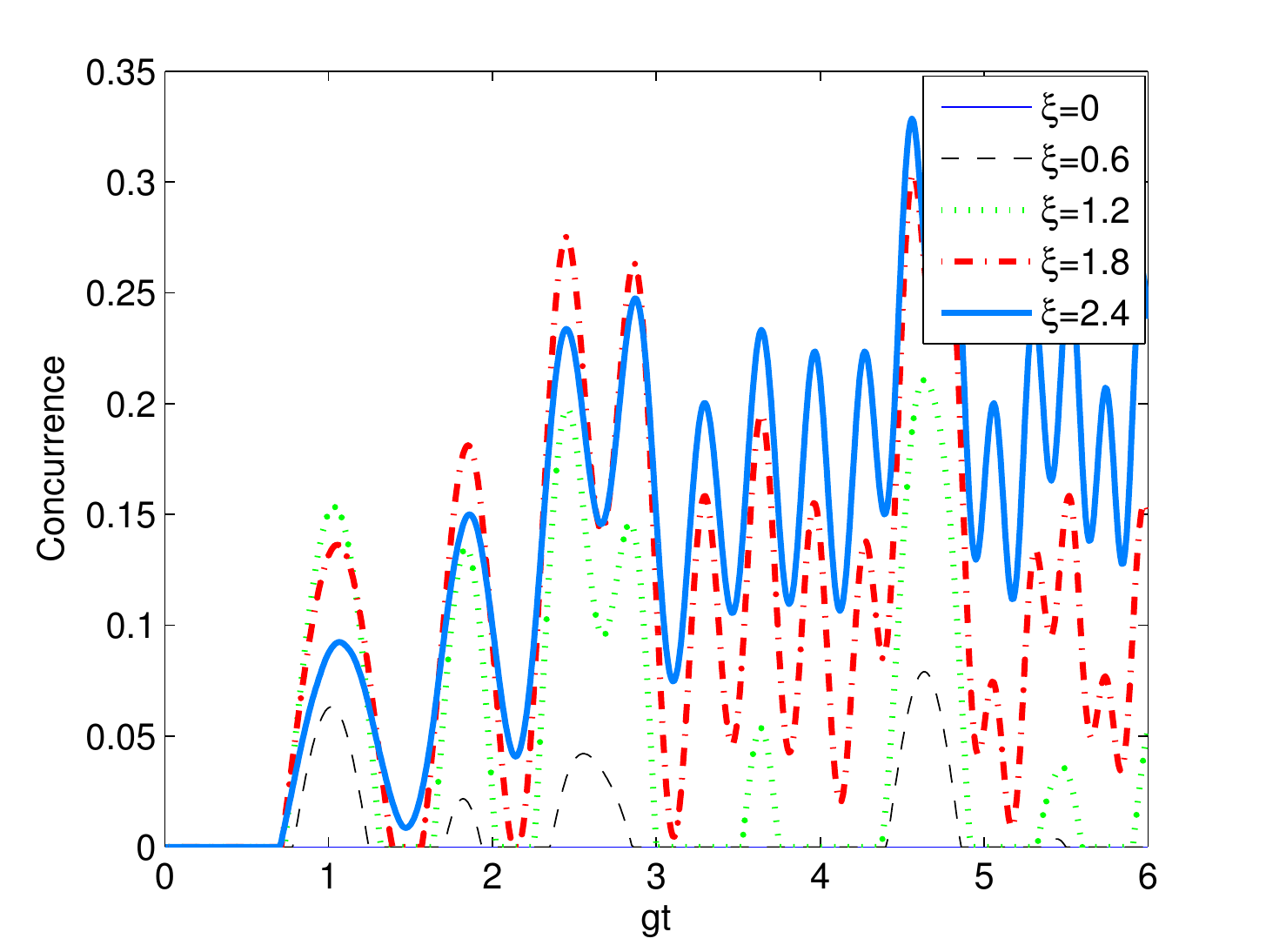}}
\subfloat[$\ket{\Phi}$ - State moves towards being maximally entangled at all times as squeezing is increased]{\label{ssqeegg}\includegraphics[width=3.2 in]{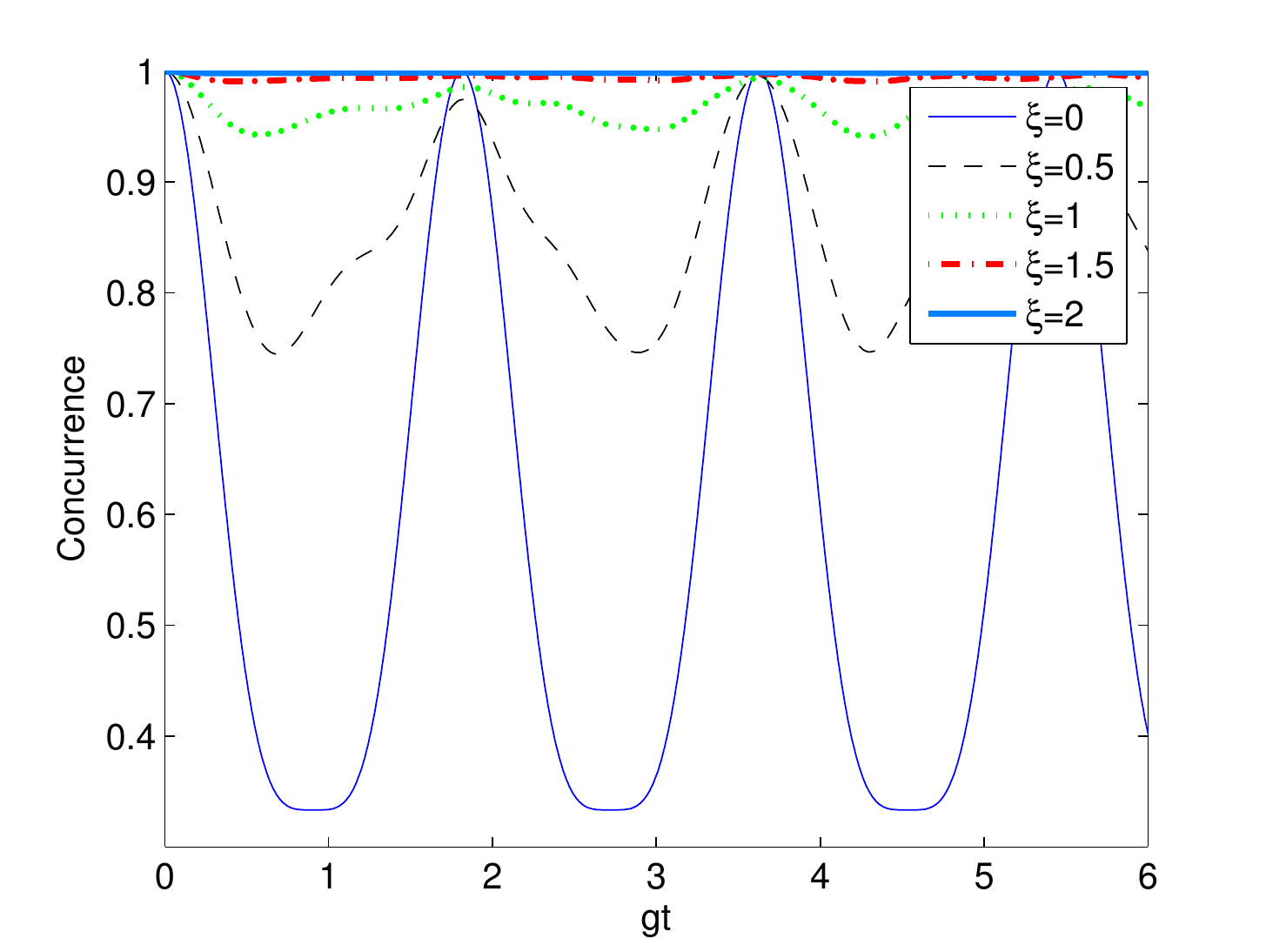}}\\
\caption{Entanglement dynamics for a single mode squeezed state $(\ket{\xi_s})$ in SMSC or two-mode squeezed state $(\ket{\xi,0,0})$ in TMSC interacting with different initial atomic states}
\label{ssq}
\end{figure}

This structure means that, as shown in Table \ref{phi0}, many familiar initial field states fail to dynamically generate entanglement in the TMAC model.  In contrast, entanglement generation is a common
feature in the TMSC model for the same selection of field states. For example, starting with a two-mode squeezed state in the field modes does not generate any entanglement in TMAC, because the corresponding field state in DJC possesses no correlations amongst $TF_{1}$ and $TF_{2}$.  On the other hand, a two-mode squeezed state in the symmetrically coupled case generates entanglement in the initially separable atomic states as can be seen in Fig.~\ref{ssq}. In fact, for initial atomic states $\ket{ee}$ and $\ket{gg}$, if the field is sufficiently squeezed then entanglement once generated sustains forever. 

\begin{figure}[ht]
\subfloat[$\ket{ee}$ - Entanglement generation from transfer of correlations from the field, optimal squeezing can be seen $\xi\approx1$ ]{\label{djctmssee}\includegraphics[width=3.2 in]{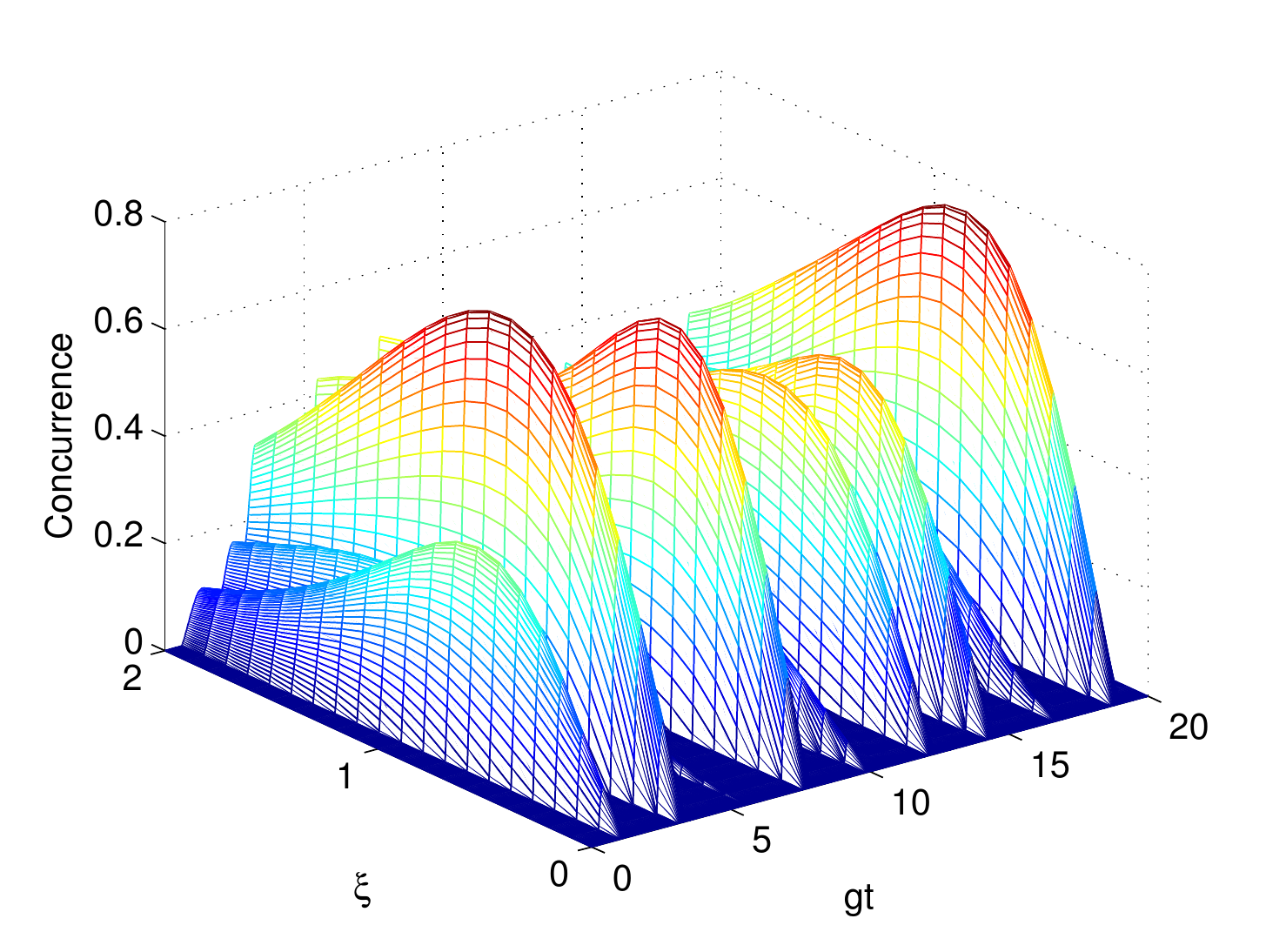}}
\subfloat[$\ket{eg}$ - Secondary peaks in entanglement vs squeezing can be observed for $\xi\approx1.5$]{\label{djectmsseg}\includegraphics[width=3.2 in]{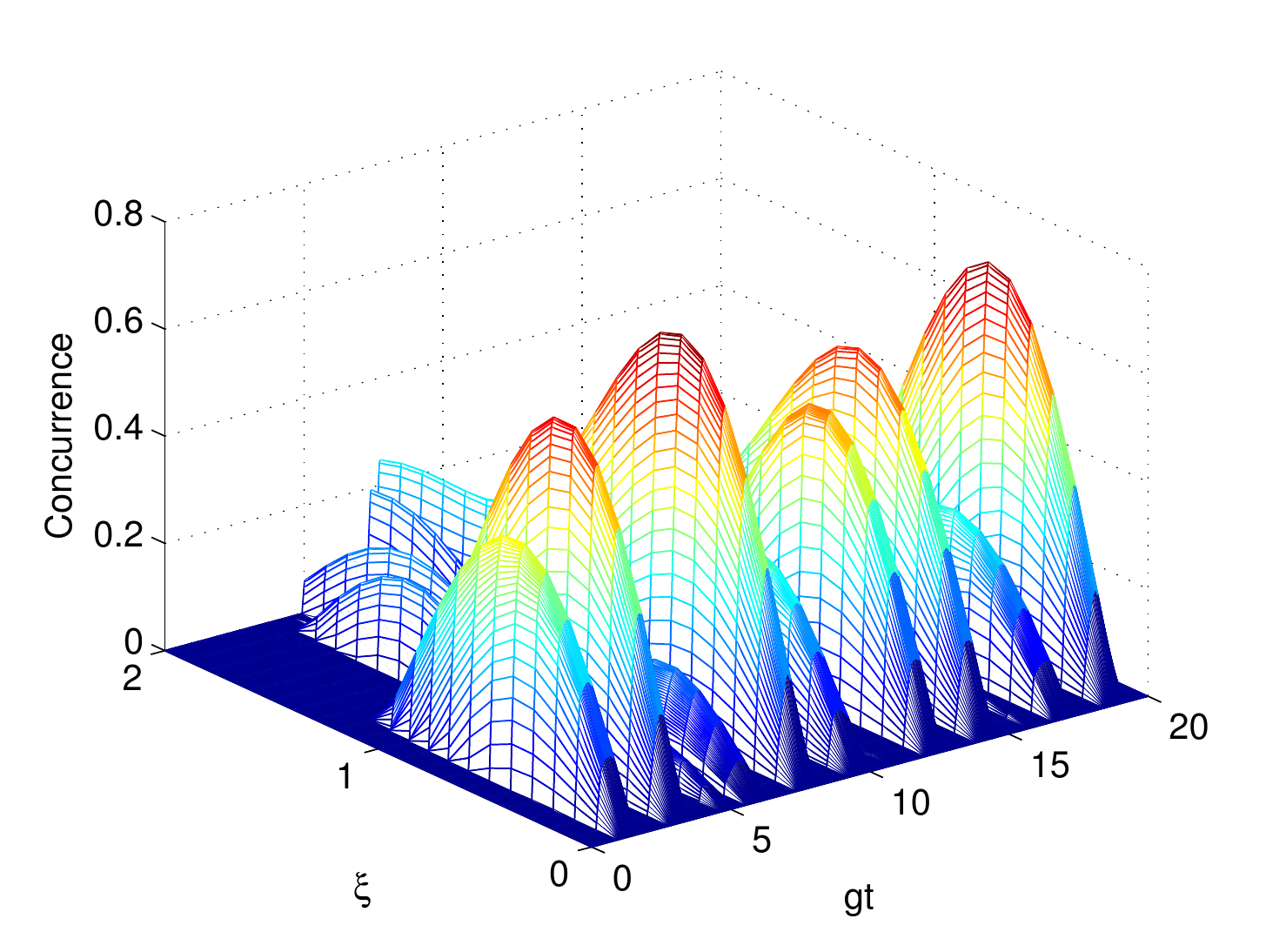}}
\caption{Entanglement dynamics for a two-mode squeezed state $(\ket{\xi,0,0})$ in DJC or product of single mode squeezed states $(\ket{\xi_s,-\xi_s,0})$ in TMAC interacting with different initial atomic states}
\label{djctmss}
\end{figure}

When considering the TMAC model with an initial field state that is a product of squeezed states $\ket{\xi_{sq},-\xi_{sq}}$ we find that entanglement is dynamically generated as shown in Fig.~\ref{djctmss}. One can understand this from the fact that the state $\ket{\xi_{sq},-\xi_{sq}}$ maps to the DJC model with an initial TMSS, so the generation of atomic entanglement occurs simply because the dynamics transfers the entanglement between the field modes to the atoms.  Further, we find that there is an optimal squeezing value that generates maximal entanglement. The maxima in generated entanglement occur for values of the squeezing parameter such that the field state is close to being a maximally entangled qubit state $(\alpha_0(\xi)\ket{00}+\alpha_1(\xi)\ket{11})$, with $\alpha_1(\xi)$ and $\alpha_0(\xi)$ being comparable. On increasing the squeezing parameter further there are contributions from higher Fock states which decrease the transfer of entanglement. We also notice some secondary peaks on increasing the squeezing parameter $\xi_{sq}$.

\begin{figure}[htp]
\includegraphics[width=4 in]{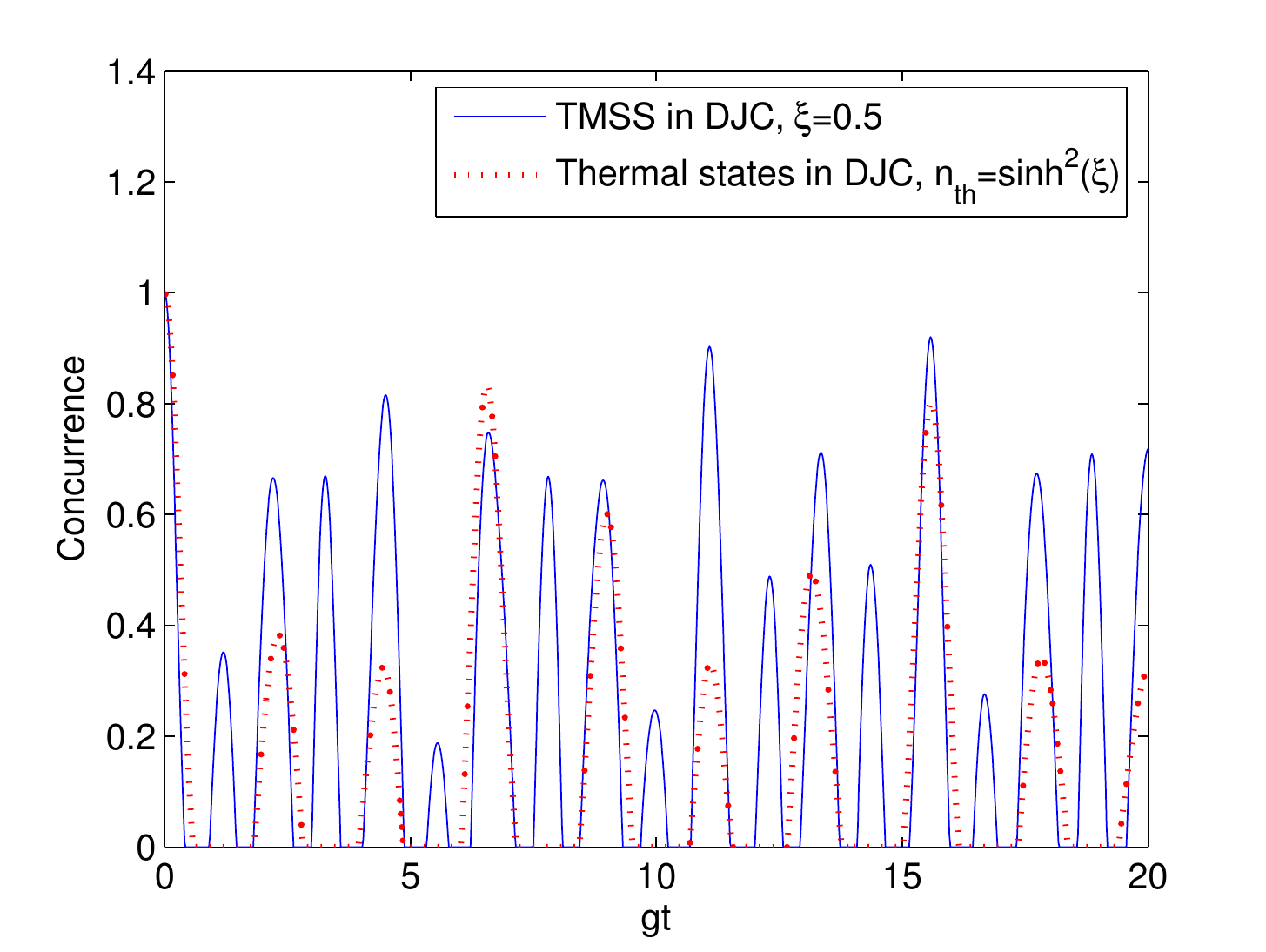}
\caption{Comparing entanglement dynamics for a two-mode squeezed state vs a thermal state in the DJC model for an initial atomic state $\ket{\Phi}$}
\label{tmsstherm}
\end{figure}

\begin{figure}[ht]
\includegraphics[width=4 in]{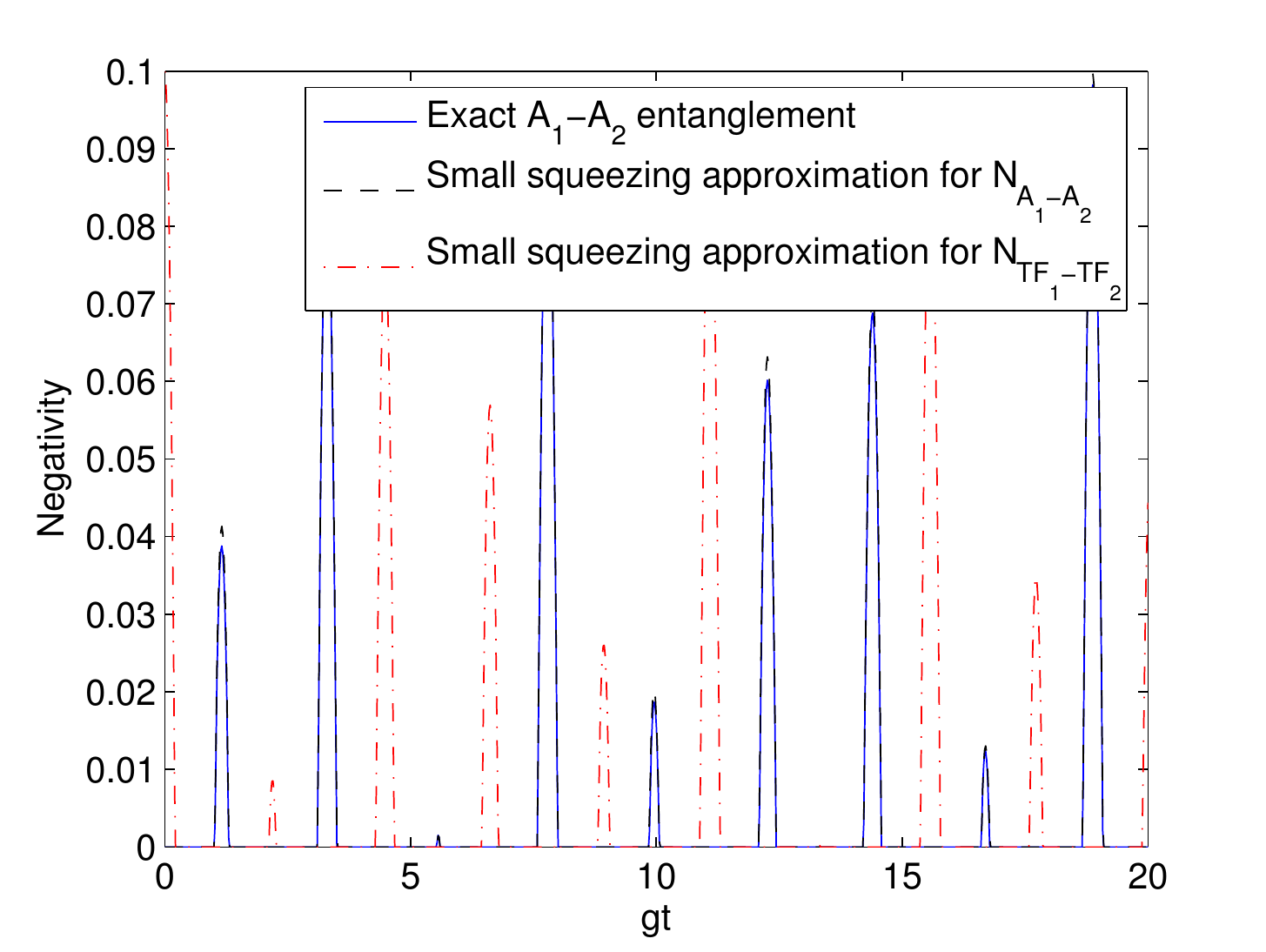}
\caption{Small squeezing entanglement dynamics for a two-mode squeezed state in the DJC model interacting with an initially separable atomic state $\ket{ee}$, entanglement is generated in the atomic subsystem via transfer of field-field correlations}
\label{lowtemp}
\end{figure}

By comparing the effects of a TMSS and thermal field state in the DJC model, 
one can also get some insight into the role of correlations between the field modes. The reduced density matrices for individual
modes correspond to a thermal state for both the cases; the difference
being that in a TMSS the two field modes are strongly correlated
with each other while in the thermal fields there are no correlations.
One would generally expect that the correlated TMSS generates more
entanglement in the atomic subsystem, which is trivially true for
the case of an initially separable atomic state. This intuition generally extends to the
situation of initially entangled atomic states, where we see that
apart from the regular SD pattern in the absence of the field-field
correlations there are spontaneously generated peaks as a result of
two-mode squeezing(Fig.~\ref{tmsstherm}); however, we note as an exception that at certain
instants of time the entanglement in the presence of an initial thermal
field exceeds that of the TMSS.
Transfer of field-field ($TF_{1}$-$TF_{2}$)
correlations can be explicitly seen in a small squeezing approximation when a two mode squeezed state interacts with an initially separable atomic state $\ket{ee}$ (Fig.~\ref{lowtemp}).
Since the probability of higher photon numbers is small, by restricting to
the 4-dimensional subspace of lowest energy states in the Fock basis for $TF_{1}$-$TF_{2}$ we obtain the negativities for the two subsystems as,
\begin{align}
\mathcal{N}_{A_{1}-A_{2}}\approx & \vert\min(s_{1}^{2}c_{1}^{2}-\xi s_{1}^{2}c_{2}^{2},0)\vert\\
\mathcal{N}_{TF_{1}-TF_{2}}\approx & \vert\min(s_{1}^{2}c_{1}^{2}-\xi c_{1}^{2}c_{2}^{2},0)\vert.\nonumber 
\end{align}
where $s_1=\sin(\sqrt{2}gt)$, $c_1=\cos(\sqrt{2}gt)$ and $c_2=\cos(2gt)$

\begin{figure}[ht]
\subfloat[$\ket{eg}$ - we observe the TMSC entanglement getting smaller as squeezing is increased, while the TMAC entanglement goes to an optimum value and then decreases]{\label{compsmsseg}\includegraphics[width=3.2 in]{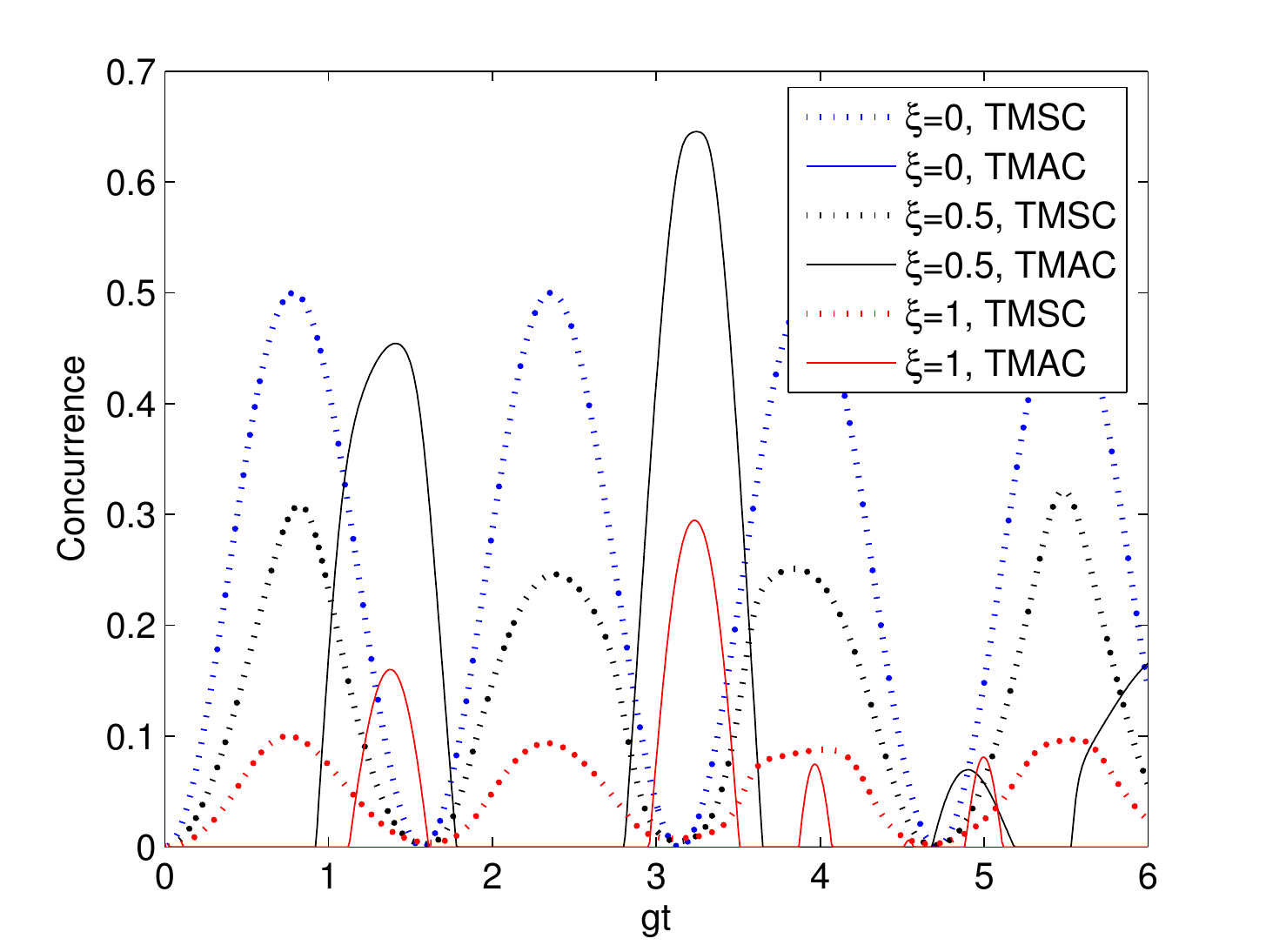}}
\subfloat[$\ket{gg}$ - TMAC shows more entanglement than TMSC, for both the entanglement increases and then decreases on increasing squeezing]{\label{compsmssgg}\includegraphics[width=3.2 in]{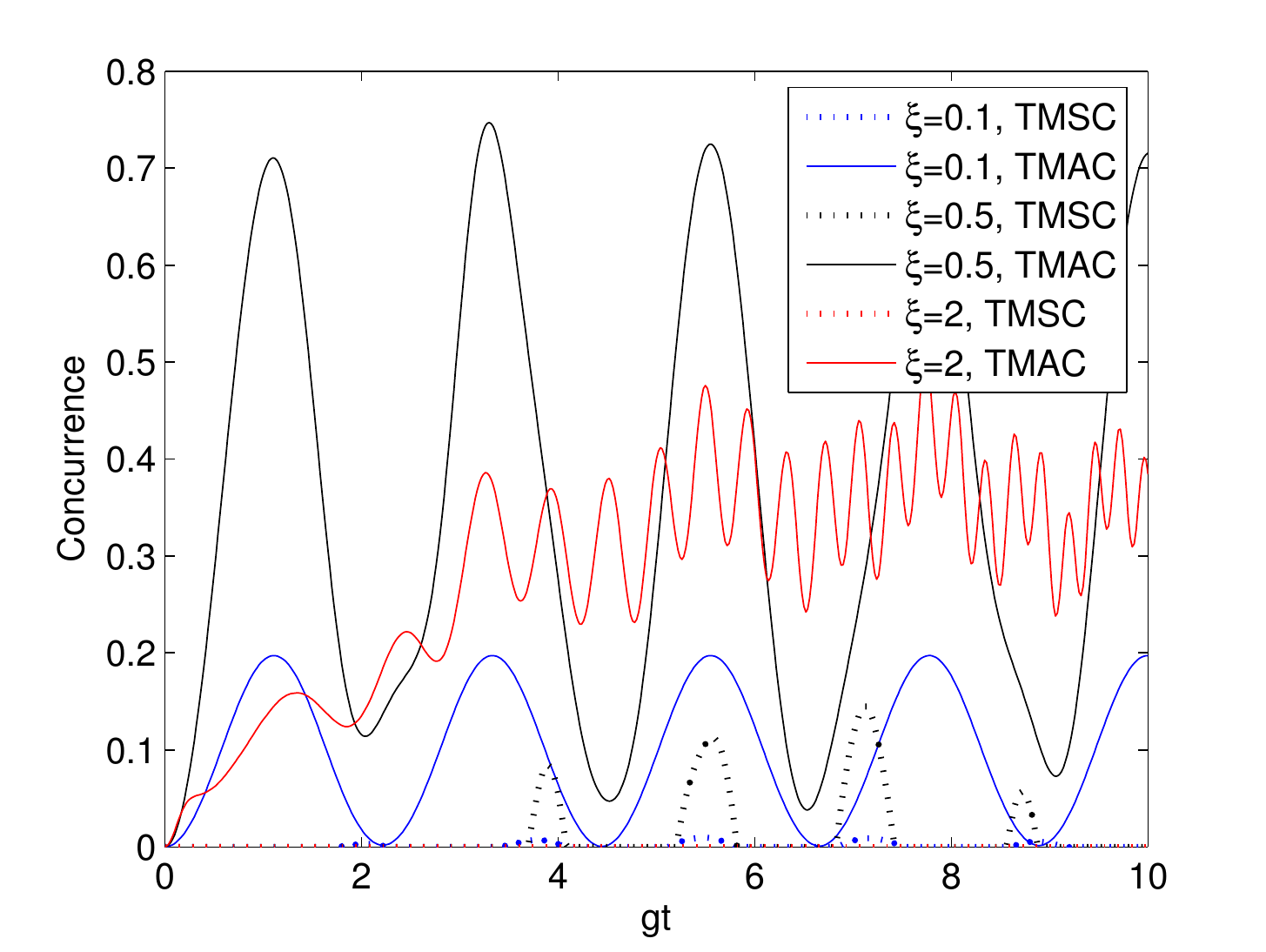}}\\
\caption{Comparison of entanglement generation in presence of product single mode squeezed states  $\ket{\xi_{sq},-\xi_{sq}}$ for TMSC and TMAC}
\label{compsmss}
\end{figure}

\begin{figure}[h!tb]
\subfloat[$\ket{eg}$ with $\ket{\eta_{nm}}$ in TMSC or $\ket{n}$ in SMSC]{\label{egn}\includegraphics[width=3.2 in]{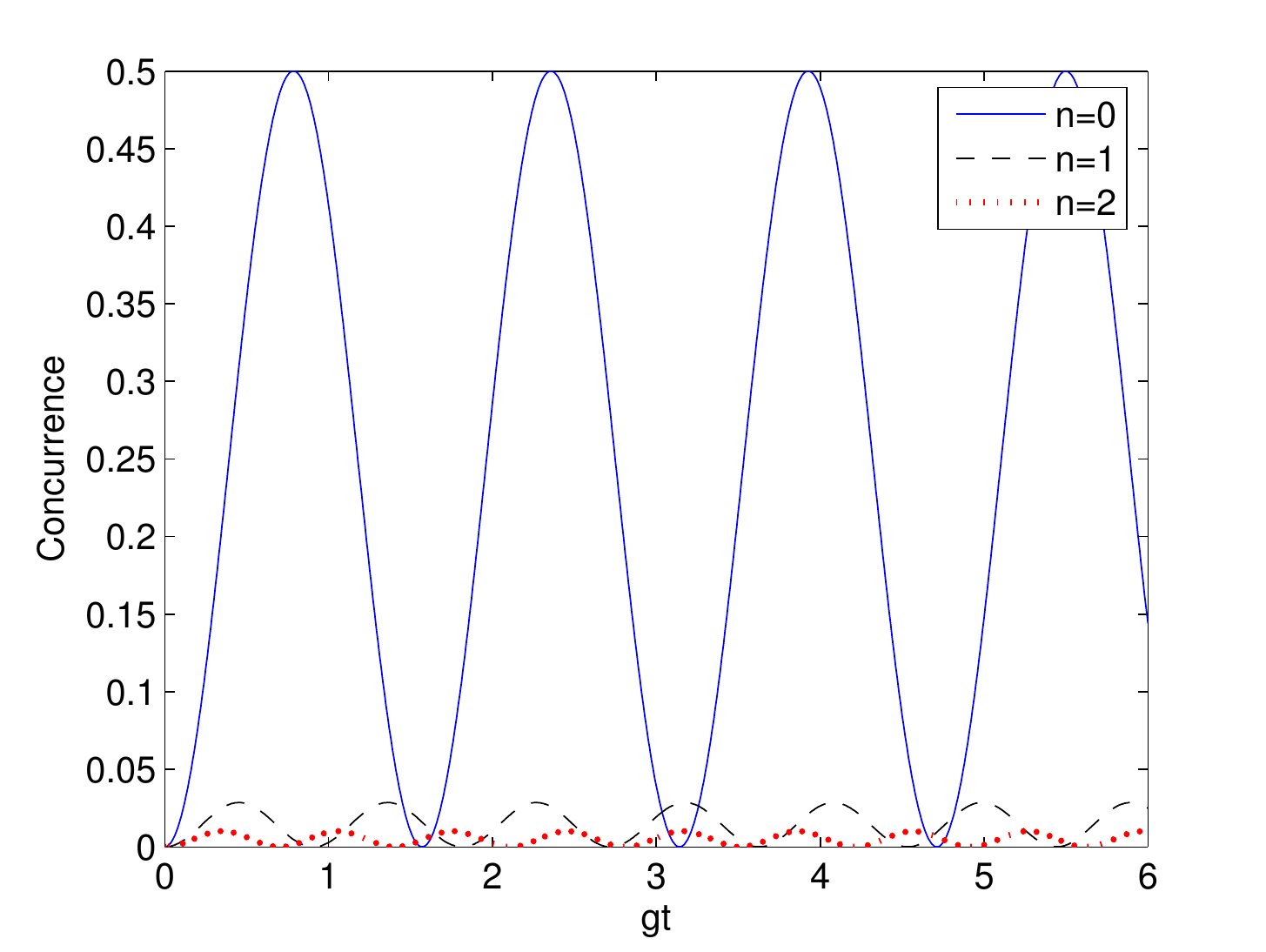}}
\subfloat[$\ket{gg}$ with $\ket{\eta_{nm}}$ in TMSC or $\ket{n}$ in SMSC]{\label{ggn}\includegraphics[width=3.2 in]{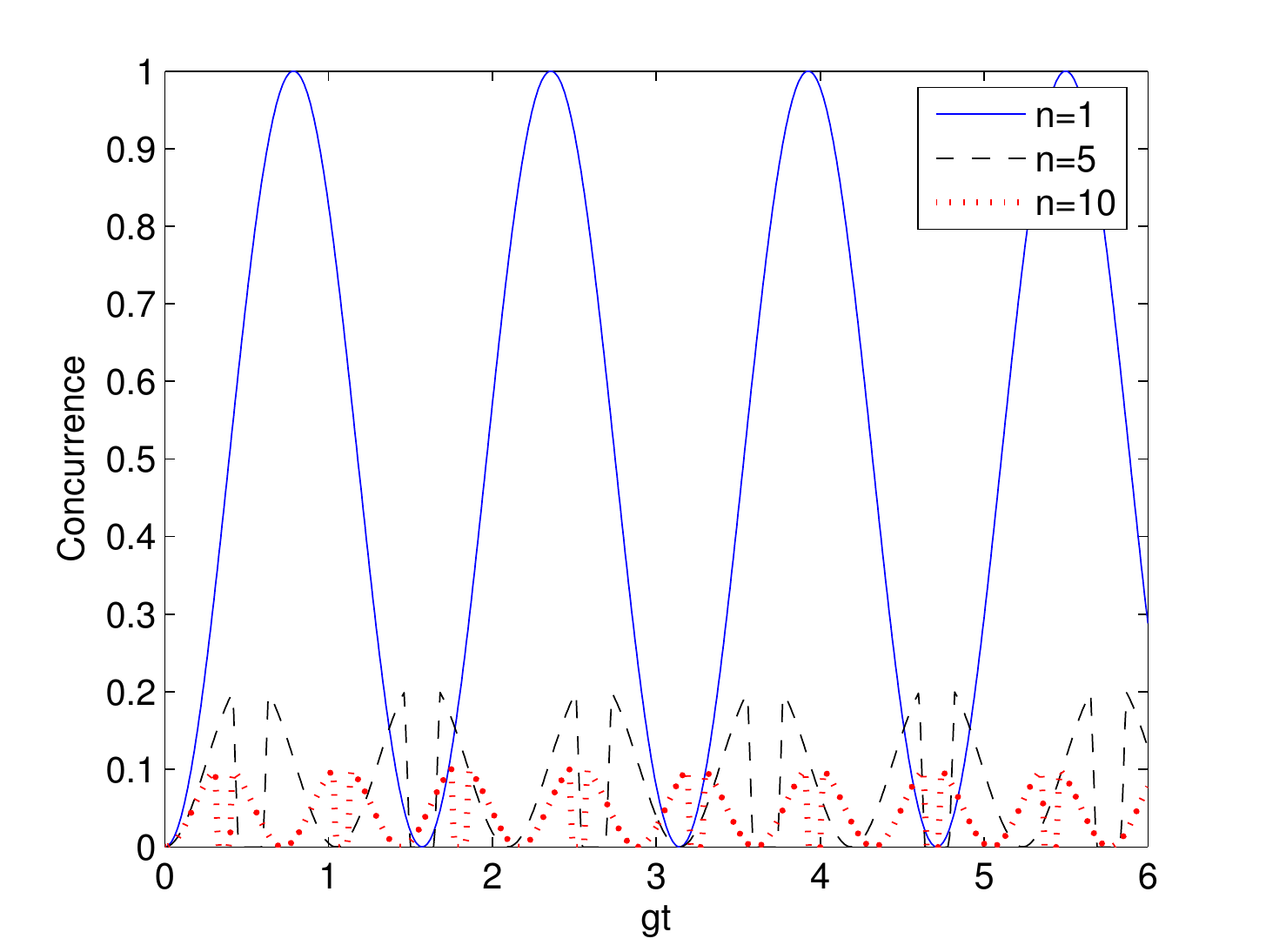}}\\
\subfloat[$\ket{eg}\bra{eg}\otimes\hat{\rho_{th}}$ in SMSC ]{\label{egth}\includegraphics[width=3.2 in]{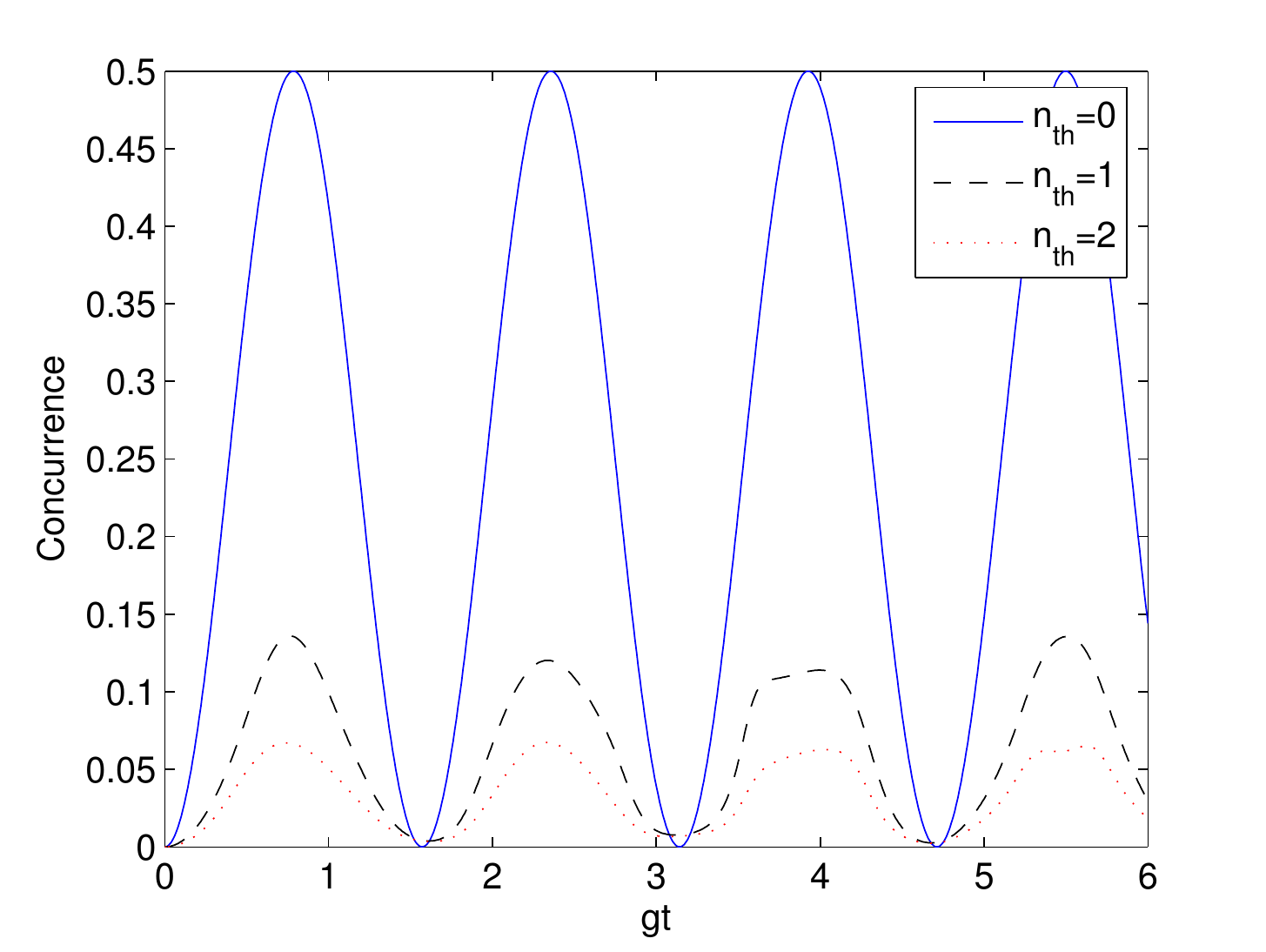}}
\subfloat[$\ket{gg}\bra{gg}\otimes\hat{\rho}_{th}$ in SMSC ]{\label{ggth}\includegraphics[width=3.2 in]{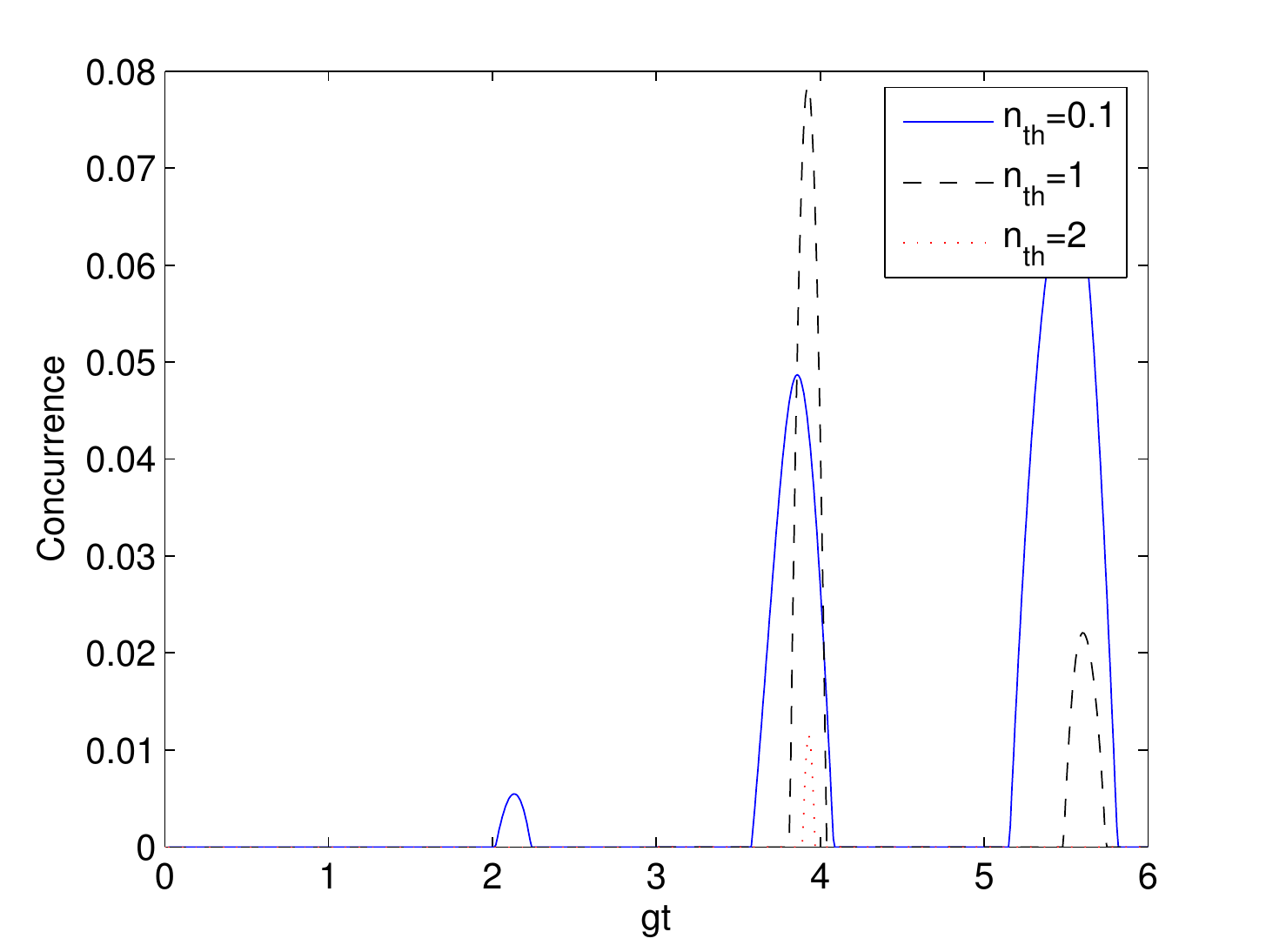}}\\
\subfloat[$\ket{eg}$ with $\ket{n_N,m_N}$ in TMSC or $\hat{\rho}_{nm}$ in SMSC  ]{\label{egnm}\includegraphics[width=3.2 in]{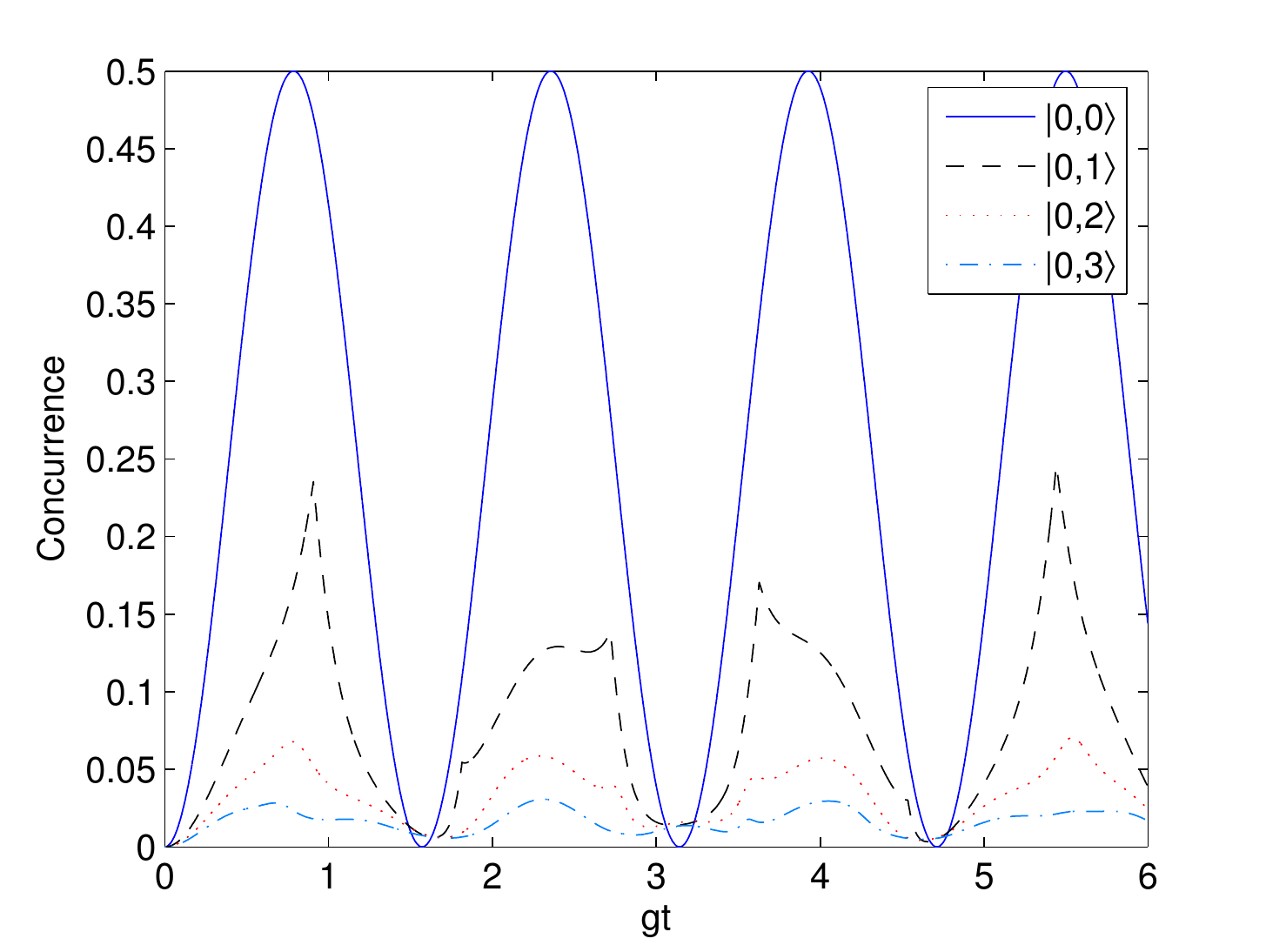}}
\subfloat[$\ket{gg}$ with $\ket{n_N,m_N}$ in TMSC or $\hat{\rho}_{nm}$ in SMSC]{\label{ggnm}\includegraphics[width=3.2 in]{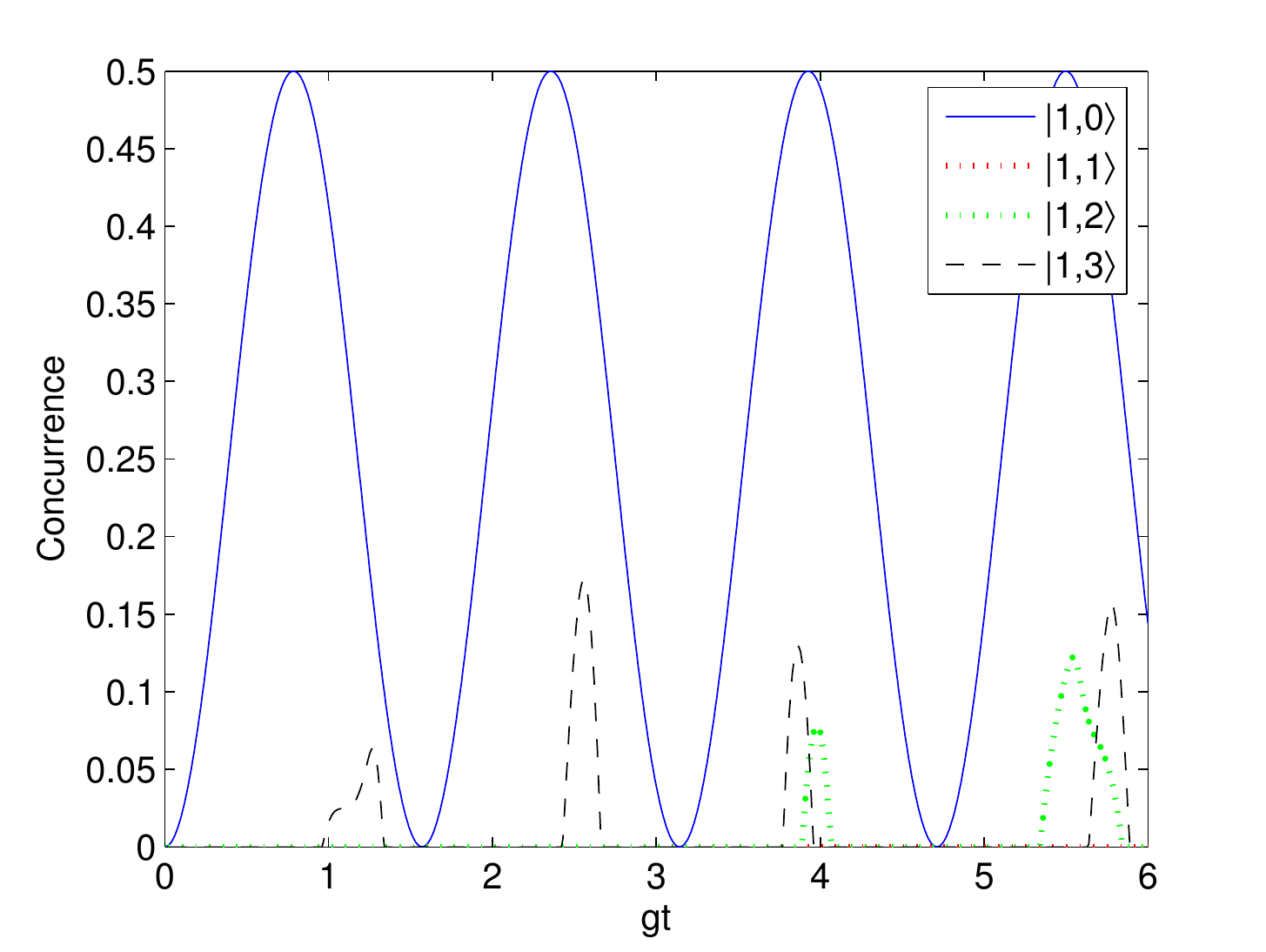}}\\
\caption{Entanglement dynamics in presence of a completely mixed initial field state in SMSC - no entanglement is seen for initial atomic state $\ket{ee}$, only DI is observed for the atomic state $\ket{eg}$ and maximum entanglement for $\bar{n}\approx1$ for $\ket{gg}$}
\label{diagonal}
\end{figure}

In the TMSC model, the same initial field state, consisting of a product of squeezed states, corresponds to having a thermal field state in the SMSC model, a situation where entanglement generation has been previously studied \cite{KimKnight2003}.  Comparing the entanglement behavior in Fig.~\ref{compsmss} for the two different couplings, we see that for the initial atomic state $\ket{gg}$ there is more entanglement generated in TMAC than TMSC.
As mentioned before, two thermal fields in TMSC also map to the same situation. This is a somewhat unintuitive situation for the generation of entanglement. There is no entanglement generation
when the atoms are initially in the state $\ket{ee}$. As Figs.~\ref{egth} and \ref{ggth} show, for $\ket{eg}$ entanglement is generated with only
instantaneous zeros, however high the temperature, and for $\ket{gg}$
there is entanglement generation and sudden death behavior with an
optimal $\bar{n}_{th}\approx1$ for the generation of atomic entanglement.

The observation that the initial atomic state $\ket{ee}$ will remain separable can, in fact, be generalized to any initial field state with a density matrix that is diagonal in the Fock basis.  As had been shown in \cite{TessierDeutsch2003}, in the SMSC model an atomic state $\ket{ee}$ interacting with any Fock state $\ket n$ in the field mode never gets entangled.\footnote{To explain this it can be observed from symmetry arguments that the time evolved atom-atom density matrix is an incoherent mixture of the states $\{\ket{ee},\ket{gg},\ket{\Psi}\}$ such that the contribution of the maximally entangled part $\ket{\Psi}$ in the mixture is at all times smaller as compared with the other two.} This then implies that the atom-atom density matrix given as $\hat{\rho}^{(n)}=\tr_{F}\left[\ket{\psi_{n}}\bra{\psi_{n}}\right]$
remains separable, where the time evolved state $\ket{\psi_{n}}\equiv\hat{U}\ket{ee}\ket n$.
Extending to a general completely-mixed field density matrix the time
evolved atom-atom density matrix given as $\hat{\rho}_{12}=\tr_{F}\left[\sum_{n}{P_{n}\hat{\rho}^{(n)}}\right]$
is clearly a convex sum of separable density matrices, and hence there
is no entanglement generation.

Looking at the initial state $\ket{eg}\ket n$ in the SMSC case, we observe no SD in Fig.~\ref{egn}. This can be explained by considering the state as a superposition $\ket{eg}=\frac{1}{2}\left(\ket{eg}+\ket{ge}\right)+\frac{1}{2}\left(\ket{eg}-\ket{ge}\right)$,
where due to the symmetry of the coupling constants the maximally entangled dark state $\left(\ket{eg}-\ket{ge}\right)/\sqrt{2}$ does
not interact with the field. It is only momentarily during the evolution that the state of the system returns to being the original separable superposition $\ket{eg}$. As a result we always observe some entanglement between the two atoms for an initial field density matrix diagonal in the Fock basis. The dynamics of the $\ket{gg}\ket n$ state in the SMSC model, shown in Fig.~\ref{ggn}, exhibit SD in general except for the special case of $n=1$ where because of symmetry reasons the state oscillates between the states $\ket{gg}\ket 1$ and $\ket{\Psi}\ket 0$, going from being separable to maximally entangled. Hence, any density matrix diagonal in the Fock basis with a high component of $\ket 1\bra 1$ would generate more entanglement in general.

Apart from the Fock state in the SMSC model with a thermal field,
another example of having a initial field density matrix diagonal
in the Fock basis is to have a Fock state $\ket{n,m}$ in the TMSC
which corresponds to the SMSC density matrix \begin{align}
\hat{\rho}_{nm}\equiv\frac{1}{2^{m+n}n!m!}\sum_{k,p=0}^{n}\sum_{l,q=0}^{m}{\kappa_{mnkl}\ket{m+n-k-l}\bra{m+n-k-l}}\end{align}
 where $\kappa_{mnkl}=\leftexp nC_{k}\leftexp nC_{p}\leftexp mC_{l}\leftexp mC_{q}\delta_{k+l,p+q}(m+n-k-l)!(k+l)!(-1)^{l}$.
For this state again we see no entanglement generation for $\ket{ee}$
and DI for $\ket{eg}$ and maximal entanglement
in $\ket{gg}\ket{10}$, as shown in Figs.~\ref{egnm} and \ref{ggnm}. Another point we observe is that for an initial
state $\ket{gg}\ket{n,n}$ there is no entanglement generation, which
is a common feature between TMSC and TMAC. A Fock state $\ket{n,m}$
in TMAC transforms into an entangled state in the DJC model, so we
expect and observe entanglement generation in the system for an initially
separable atomic state. As an exception we find that for $n=m$, if
there is no atom-atom entanglement to begin with then the atoms remain
separable evolving into a completely mixed state. This is counterintuitive in the sense that even though the field state in the DJC model is entangled to begin with, there is still no transfer to the atomic subsystem. This feature can be explained by considering the initial field state $\ket{\eta_{nn}}=\frac{1}{2^n n!}\sum_{k=0}^n{\leftexp{n}C_k(-1)^k\sqrt{2k!(2n-2k)!}\ket{2n-2k,2k}}$. If the atoms are initially in the state $\ket{ee}$ then time evolution will lead to an entanglement of atomic and field states such that detecting whether the number of photons in each mode is even or odd tells us the state of the two atoms.  On tracing out the field this gives us a completely mixed atom-atom density matrix with no atom-atom entanglement.  The same is true for the initial atomic states $\ket{eg}$ and $\ket{gg}$.

While there are more nuanced details in what we have reported in this section, generically speaking it seems that the TMSC (and SMSC) models are much more conducive to the dynamical generation of entanglement than the TMAC (and DJC) model, suggesting that atomic separation may have a strong influence on this.  Particularly useful is the ability in the TMSC to dynamically generate entanglement with a thermal field state or from a squeezed state in SMSC where one can even generate atomic entanglement that is AL.

\section{Entanglement Sudden Death and Protection\label{sec:EP}}

The phenomenon of entanglement sudden death has clearly provoked much
theoretical interest, and it is related to another question that is
both interesting from a theoretical perspective and clearly of great
practical importance: how can one protect a system from disentanglement?
Here we do not propose any active scheme for protecting entanglement
(as in, e.g. \cite{ManiscalcoEtAl2008}), but rather consider what
initial states of the field tend to minimize the loss of entanglement
or safeguard entanglement once it has been dynamically generated.
Of particular interest is avoiding SD.

\begin{figure}[htp]
\includegraphics[width=4 in]{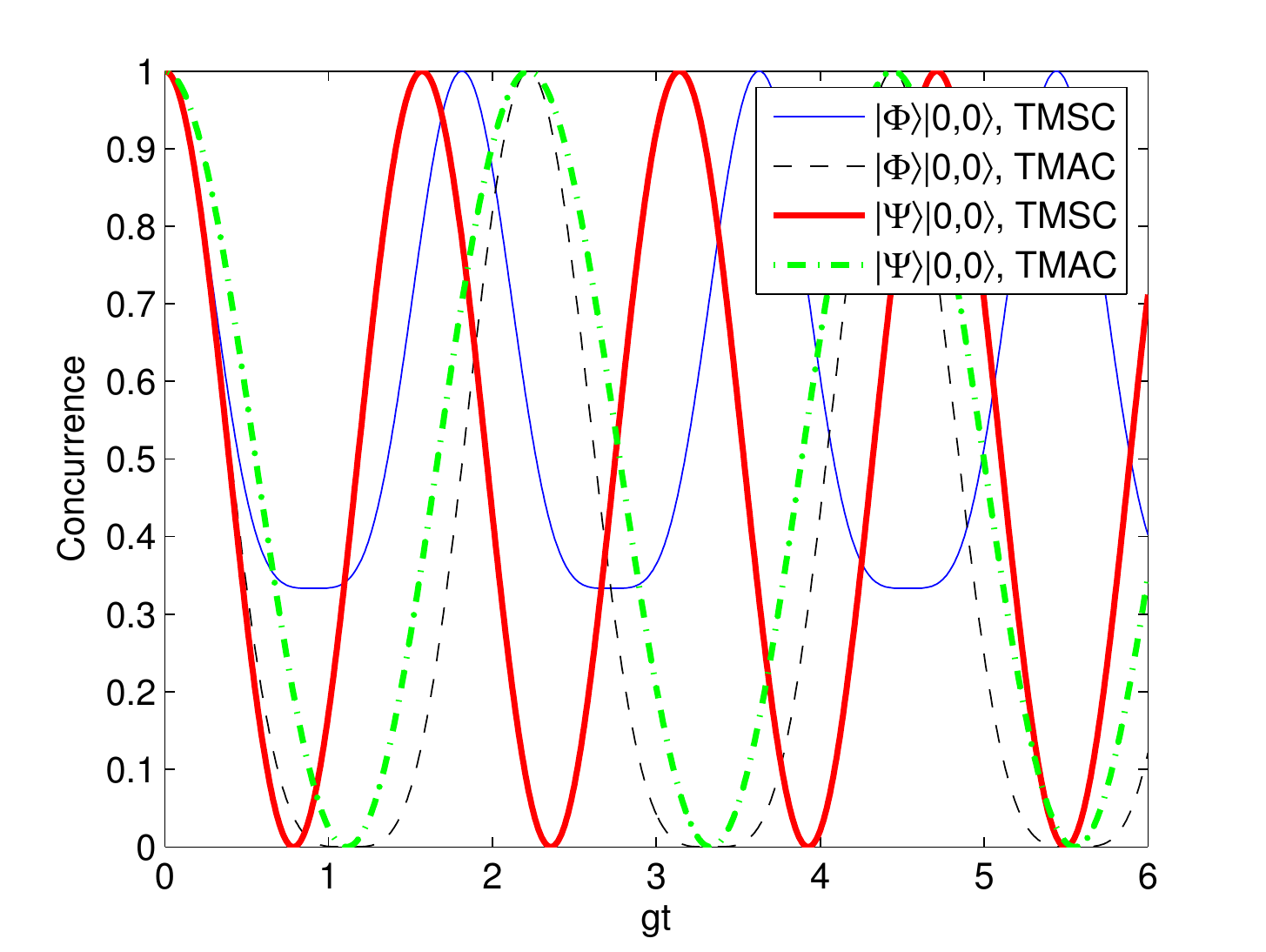}
\caption{Entanglement dynamics in vacuum interacting with initially entangled atomic states $\ket{\Phi}$ and $\ket{\Psi}$ in presence of symmetric and anti-symmetric couplings}
\label{eeggegge}
\end{figure}

\begin{figure}[ht]
\includegraphics[width=4 in]{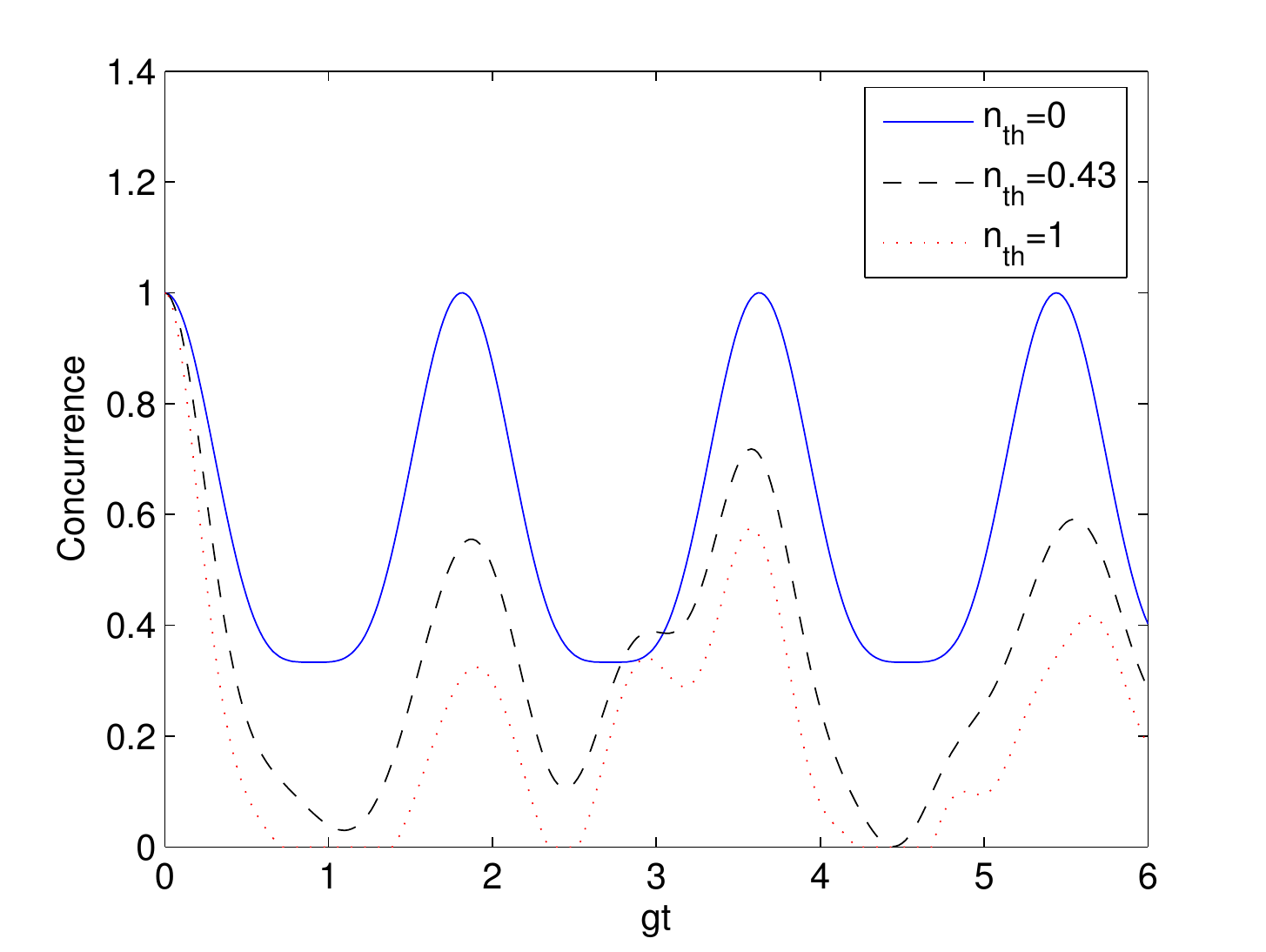}
\caption{Entanglement dynamics for a single mode thermal field interacting with an initially entangled atomic state $\ket{\Phi}$ in the TMSC model. SD occurs after a certain threshold temperature.}
\label{singlethermeegg}
\end{figure}

In terms of the effect of the initial atomic state on the entanglement
dynamics, it has been discussed previously in \cite{YonacYuEberly2007}
for the DJC model that the initial state $\ket{\Phi}\ket{00}$ undergoes
sudden death while $\ket{\Psi}\ket{00}$ has DI
of the atomic entanglement. This differentiation in behavior is common
to many of the models for studying SD in which each atom interacts with a separate field \cite{YuEberly2004,YuEberly2009}.
The same situation must occur in the TMAC case as well. In comparing
these two initial states for the TMSC we find a sort of reversal of roles; as Fig.~\ref{eeggegge}
shows, $\ket{\Phi}\ket{00}$ enjoys non-zero entanglement at all times
and $\ket{\Psi}\ket{00}$ still suffers DI, so now the entanglement of $\ket{\Phi}$ is better preserved. 
This reversal also holds in the case of a thermal field in the TMSC model so long as the thermal average photon number is below a threshold value $\bar{n}_{crit}\approx0.43$ (Fig.~\ref{singlethermeegg}) with the state being AL (above this critical temperature $\ket{\Phi}$ experiences SD as well).

\begin{figure}[ht]
\subfloat[$\ket{\xi,0,0}\ket{\Phi}$ - SD for TMAC, entanglement moves towards maximal entanglement in TMSC]{\label{compeeggtmss}\includegraphics[width=3.2 in]{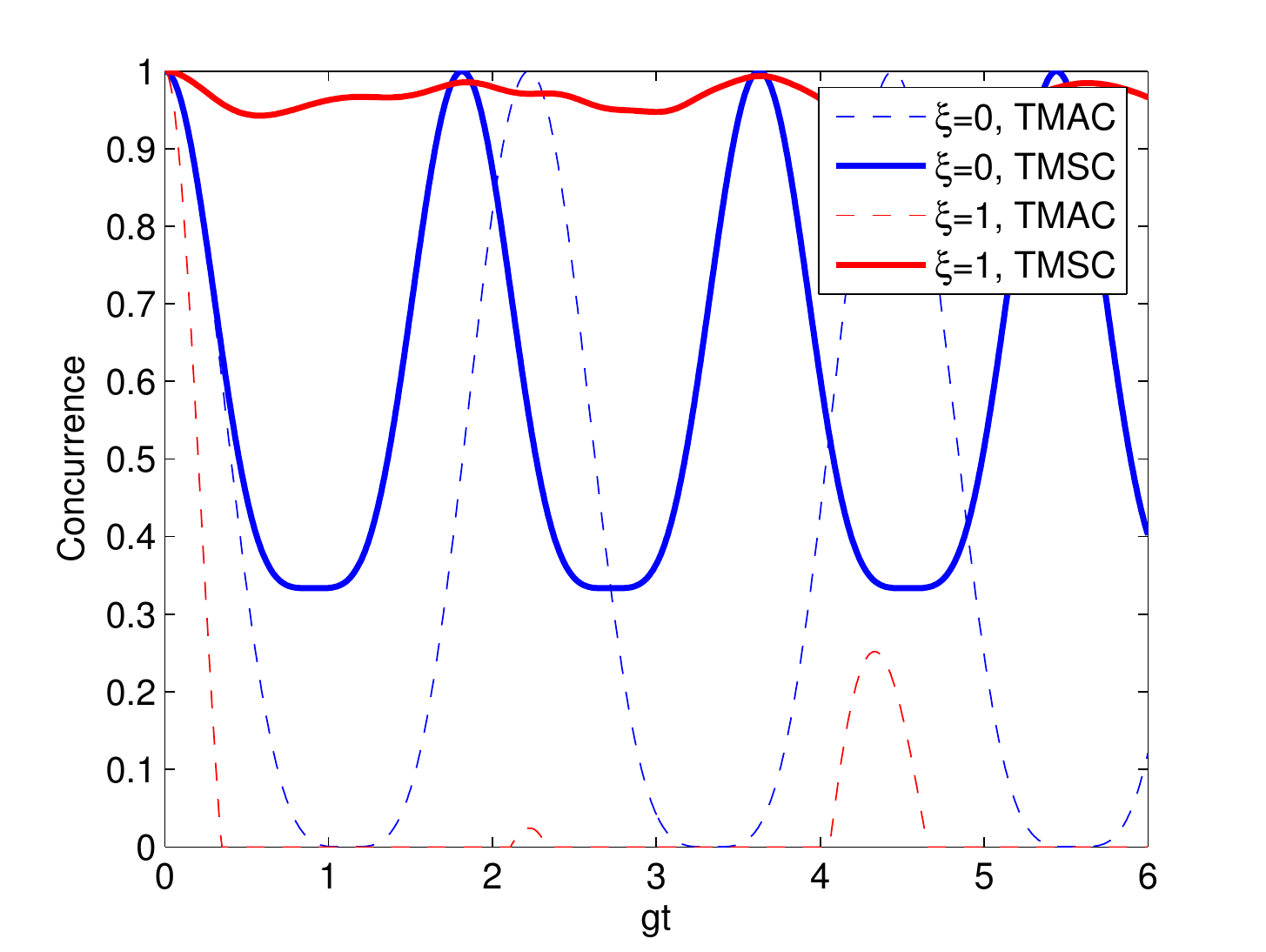}}
\subfloat[$\ket{\xi,0,0}\ket{\Psi}$ - SD for both TMSC and TMAC]{\label{compeggetmss}\includegraphics[width=3.2 in]{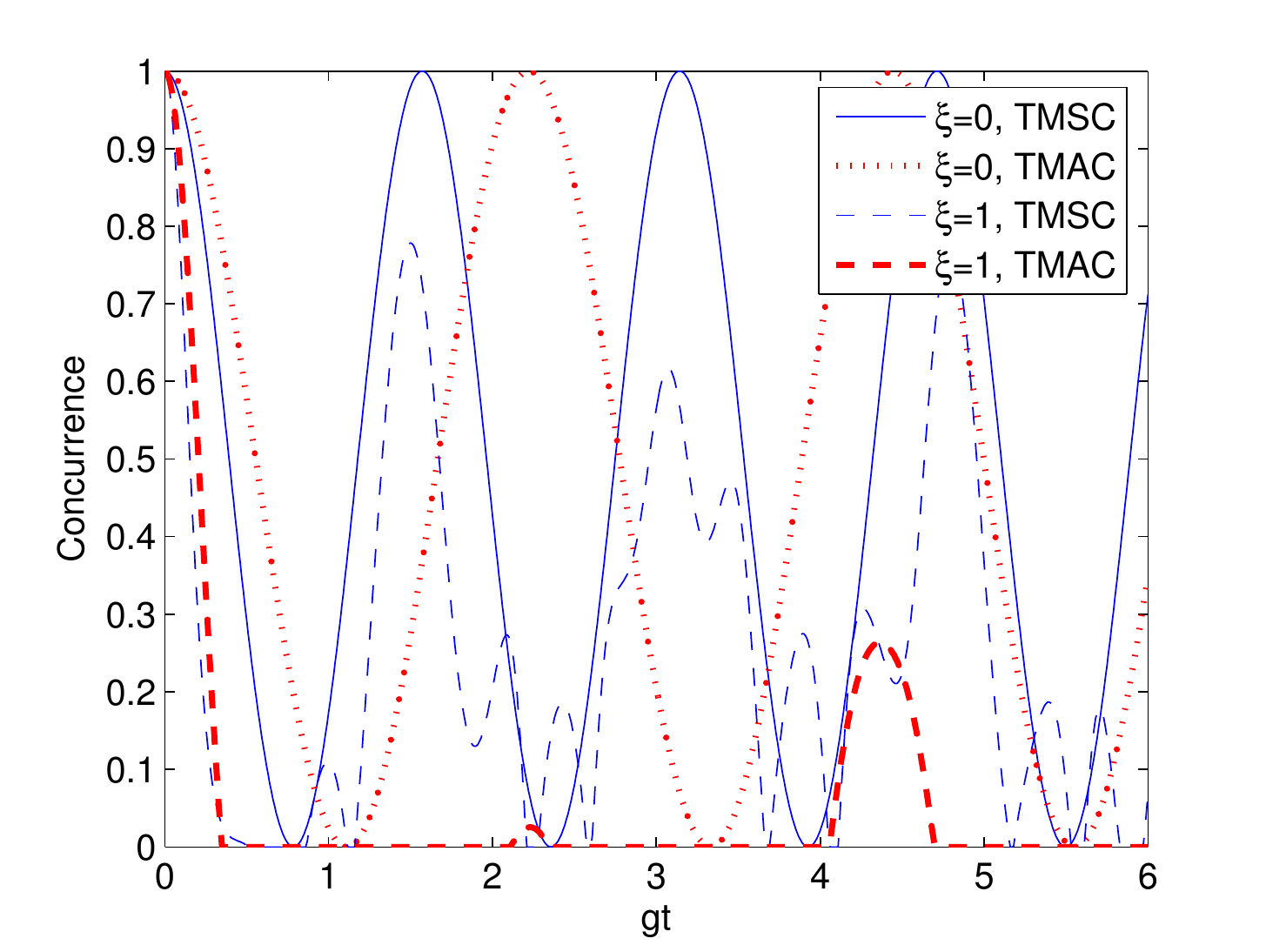}}\\
\caption{Comparing entanglement dynamics for the initially entangled atomic states interacting with a two-mode squeezed state  and  in TMSC and TMAC}
\label{comptmss}
\end{figure}

If one considers the TMSC model with the field modes in a two-mode squeezed state, then for the separable initial atomic states $\ket{ee}$ and $\ket{gg}$, if the field is sufficiently squeezed, entanglement is dynamically
generated and once generated sustains forever (whereas a Fock
state or thermal state may generate entanglement but it goes to zero again at some later time).  This behavior is shown in Figs.~\ref{ssqee} and \ref{ssqgg}.  If the atoms are initially in the entangled atomic state $\ket{\Phi}$, then Fig.~\ref{comptmss} shows that in the TMSC model increasing squeezing raises the minimum value of entanglement
progressively towards a situation where the state is maximally entangled at all times, while
for the TMAC model the system exhibits SD, and the entanglement gets destroyed in general with increasing squeezing.

\begin{figure}[ht]
\subfloat[$\ket{ee}$ - No sudden death for a range of values of $\alpha$]{\label{cohee}\includegraphics[width=3.2 in]{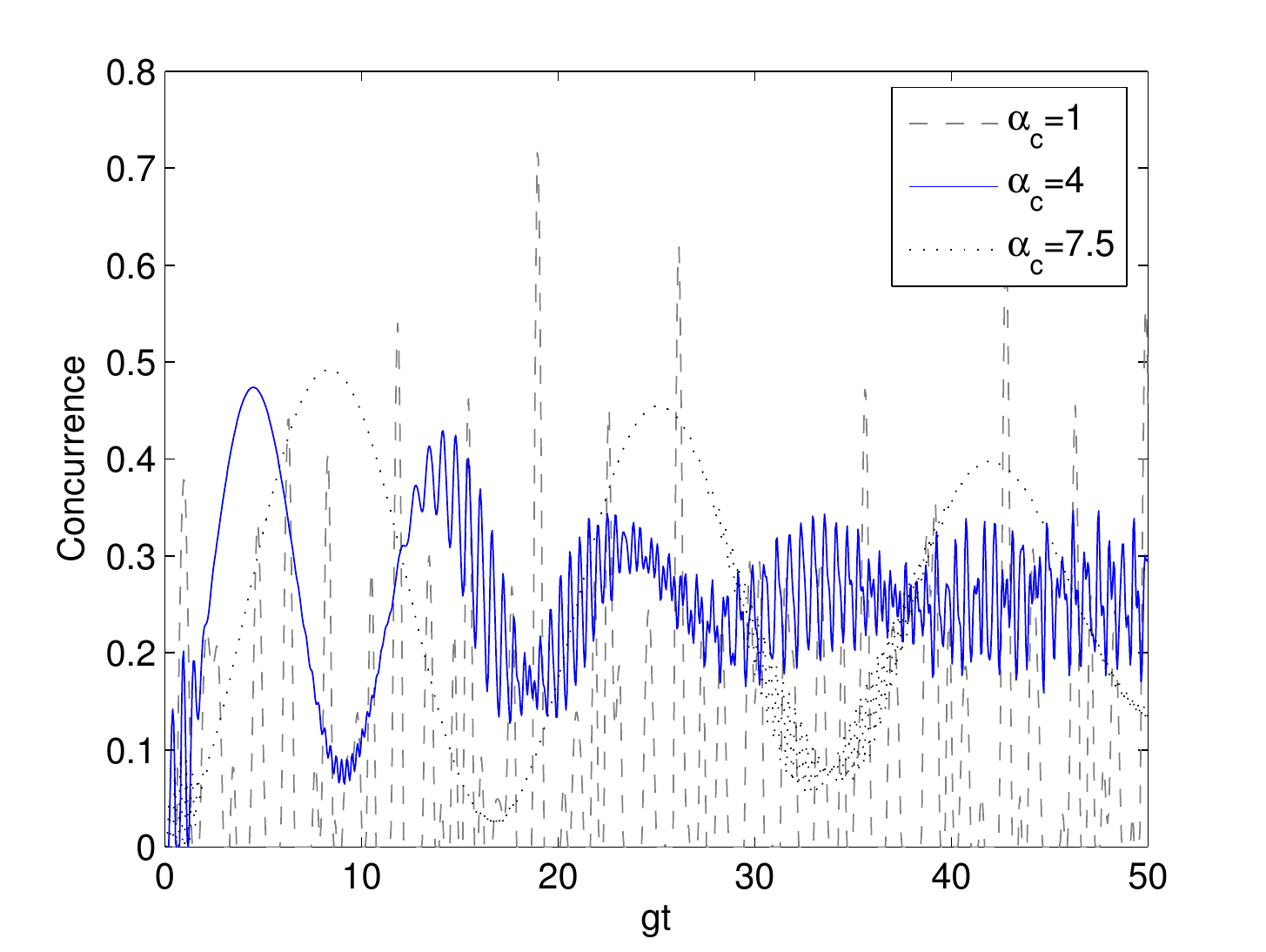}}
\subfloat[$\ket{eg}$ - Sudden death for all $\left\vert\alpha\right\vert>0$]{\label{coheg}\includegraphics[width=3.2 in]{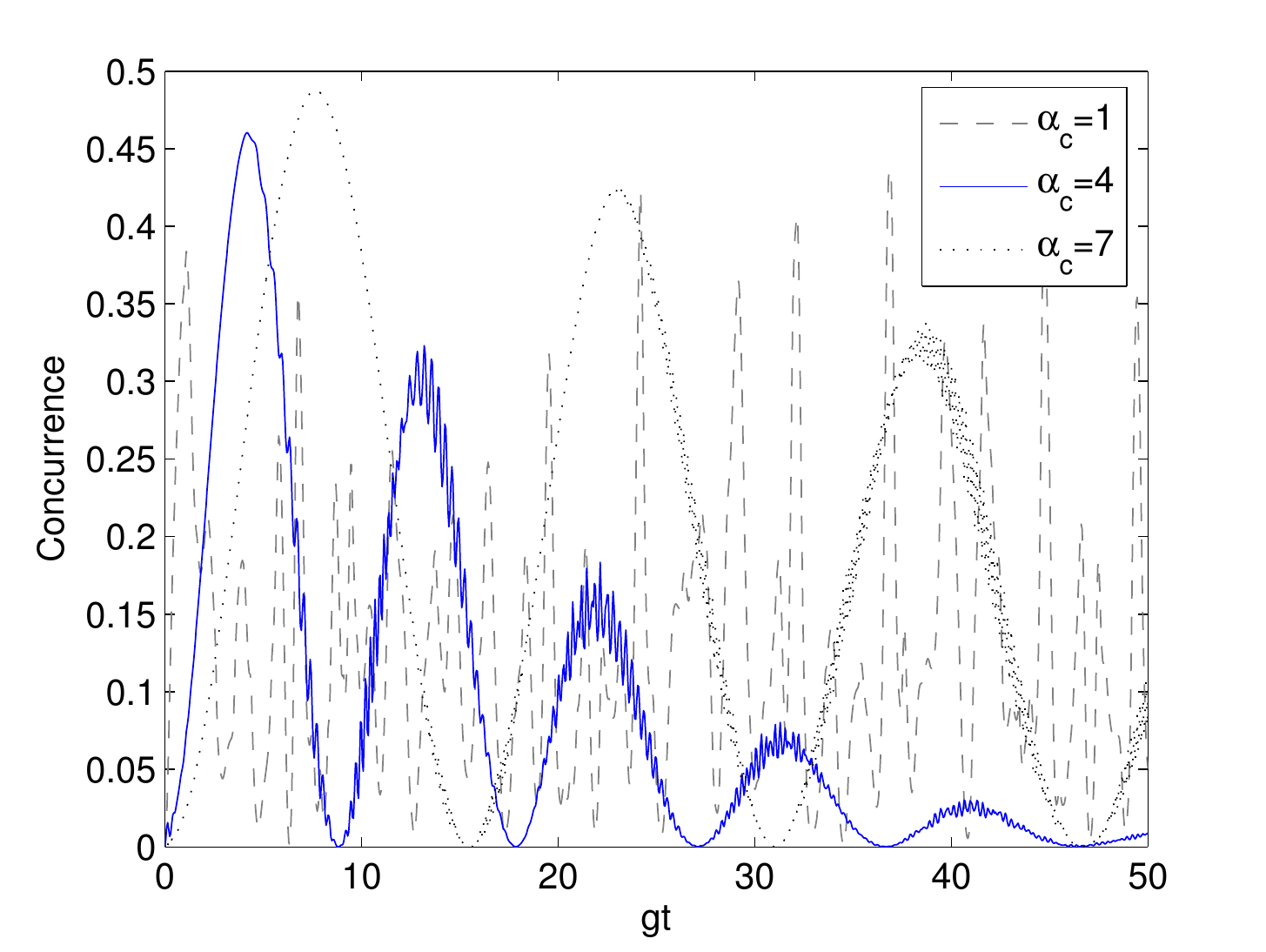}}\\
\caption{Entanglement generation in presence of a coherent state in SMSC}
\label{coh}
\end{figure}

A general observation that seems to fit for most of the cases considered
is that a higher average number of photons in the field destroys entanglement. As an exception, however, it can be seen from Fig.~\ref{cohee} that in the TMSC model with the field in a product of coherent states, if the average number of photons is in a particular range then there is no SD once entanglement is generated for the states $\ket{ee}$ and $\ket{gg}$. While Fig.~\ref{coheg} shows that for $\ket{eg}$, the regular rule applies. In the case of the initial field being a two-mode squeezed state (Fig.~\ref{ssq}), we observe that after a threshold squeezing (or average photon number) there is no sudden death of entanglement after generation in the system when the atoms are initially in the state $\ket{ee}$ or $\ket{gg}$. On the other hand, for the initial atomic state $\ket{eg}$ entanglement decreases as the average number of photons is increased.

Much as for generating entanglement, we find that the TMSC model is generically better suited to protecting entanglement from sudden death.  While in the TMAC model at nonzero temperature disentanglement occurs for both entangled states considered, in the TMSC we find that below a threshold temperature entanglement of the $\ket{\Phi}$ state remains AL.  In addition, we find that in the SMSC model a two-mode squeezed vacuum field can keep $\ket{\Phi}$ almost maximally entangled at all times, provided squeezing is large enough.  In the TMAC sudden death is a quite generic feature, which is only escaped for an initially entangled state when the field is the vacuum or a select product Fock state.

\section{Summary and Discussion\label{sec:conclusion}}

We have analyzed the entanglement dynamics in a model consisting of
two  two-level atoms and two electromagnetic field modes for a
variety of familiar field states and classified the various cases in
terms of phenomena such as dynamical entanglement generation,
entanglement sudden death.  One aim of this analysis is to get a sense of the variety of different classes of entanglement dynamics that can arise from different atomic separations in a case where two atoms interact with a shared EM field.  It is useful to examine this question in a very simple model that can be solved with a minimum of approximations (using which can
lead to unphysical effects) and understood in detail. We have argued
that there is no non-trivial distance dependence in a single mode
model, and, therefore, a two-mode model represents the simplest case
for the study of distance dependence in the entanglement dynamics.

We have studied two arguably extreme cases out of the class of
two-mode Hamiltonians that can arise, one in which the two modes are
symmetrically coupled (TMSC) to the second atom, and one where the
two modes have asymmetric coupling (TMAC).  A useful insight in
understanding these models is that the atomic dynamics in the TMSC
model correspond exactly to the dynamics for a model with single
field mode symmetrically coupled (SMSC) to both atoms with a suitable
mapping of the field state, while the atomic dynamics for the TMAC
model correspond exactly to the dynamics for a double Jaynes-Cummings
(DJC) model, made up of two isolated subsystems each with one atom
coupled to one mode, under the proper transformation of the field
state. These mappings help one understand the significant differences
in behavior in the two seemingly similar models, giving a window into
how significantly atomic separation can affect entanglement dynamics.
Another significant implication of the mapping between the TMSC and
SMSC models comes from the fact that the mapping of initial field
states between those models is many-to-one (because it involves a
partial trace); this shows us entire classes of field states for the
TMSC model that will give exactly the same atomic dynamics. In
particular, we saw that a product squeezed state can be identical to
a thermal state with respect to the atomic dynamics.

In examining the dynamical generation of entanglement  from an
initially separable atomic state in Sec.~\ref{sec:EG}, we find quite
a marked contrast between the TMSC and TMAC models.  While
entanglement generation is a relatively common feature, present for a
variety of field states, in the TMSC model, it is comparatively much
more rare in the TMAC model.  One aspect that highlights these differences is that in the TMAC model entanglement can be generated by a product of squeezed states or fock states, but in the TMSC model it can also be generated by more easily prepared states including thermal states and product coherent states. This difference is not so surprising
however, if one views it in terms of the mappings we have introduced
to the other models.  When considering that the TMAC maps to the DJC
model, where the two subsystems are isolated, one would expect
entanglement generation to be relatively rare; it can only exist in
cases where the field state is mapped to an entangled state whose
entanglement can then be transferred to the atoms.  With the TMSC, by
contrast, we have a mapping to the SMSC model where a single shared
field mode can readily introduce entanglement between the two atoms.

When one is concerned with protecting the entanglement of two initially entangled atoms, our analysis in Sec.~\ref{sec:EP} shows that again the TMSC model is better for that purpose in most cases.  In the TMAC entanglement sudden death (SD) is a fairly generic feature, with the initial state $\ket{\Psi}\ket{00}$ being one of only two classes we consider that does not show SD; the $\ket{\Psi}\ket{00}$ shows "death for an instant" (DI), where entanglement goes to zero only on a finite set of points.  The TMSC shows a reversal of the fortunes of the $\ket{\Psi}\ket{00}$ and $\ket{\Phi}\ket{00}$ states, with the former experiencing DI while entanglement for the latter is AL, staying non-zero for all times.  Moreover, the DI property of the dynamics of the $\ket{\Psi}\ket{00}$ state in the TMAC model is fragile, in the sense that it is destroyed by even the smallest departure from the vacuum to, for example, a finite temperature field, while for the TMSC model the AL feature of the dynamics of the $\ket{\Phi}\ket{00}$ state is robust, remaining for non-zero temperature below a threshold $\bar{n}_{crit} \simeq 0.43$. This gives us a condition for protecting entanglement in this case.

A different sort of issue we have touched on briefly is the role of quantum correlations between field modes in the entanglement dynamics of the atoms.  In the DJC model we compared the resulting entanglement dynamics for fields in a two-mode squeezed state (TMSS) and a thermal state.  For appropriate choice of the squeezing parameter, the TMSS has the same reduced density matrix for either mode alone as the thermal state, so in this sense these form a pair of extreme cases to compare, one with strong quantum correlations while the other has none at all.  It is necessarily true in the DJC that a thermal state cannot generate entanglement, while a TMSS does.  Intuition would also suggest that the local entropy destroys entanglement and leads to qualitative sudden death features, while the entanglement generation we observe in the case of a two-mode squeezed state arises from the field-field entanglement being transferred to the atomic subsystem. The transfer of entanglement from field-field to atoms was analytically verified for a small squeezing approximation of the two-mode squeezed state where one can observe the field-field entanglement going to zero as the atom-atom entanglement builds up.  For initial entangled states we would expect that the correlations of the TMSS aid in maintaining the entanglement of an initially entangled state, and we find this to generally be the case, although in some cases for brief periods of time a thermal state can actually result in greater atomic entanglement than the corresponding squeezed states.

As a general result for the class of density matrices that are completely mixed in the Fock basis in SMSC case, we conclude that in terms of generating entanglement it is preferable to choose an initial atomic state $\ket{eg}$, which has entanglement being AL or at the least DI, as opposed to $\ket{ee}$ where no entanglement is generated. This had been previously pointed out in the case of Fock states and thermal states \cite{KimKnight2003,TessierDeutsch2003}.

As an interesting result in terms of entanglement protection, we find that a single mode squeezed state interacting with symmetrically coupled atoms initially in the state $\ket{\Phi}$ can be extremely effective in protecting the entanglement.  Even for the vacuum, this entangled state is AL, but as squeezing is increased the minimum entanglement rises monotonically toward maximal entanglement.  For entanglement generation by a single mode squeezed state in the separable atomic initial states $\ket{ee}$ and $\ket{gg}$, we found that for a sufficient amount of squeezing entanglement is not only generated but remains AL for all future times. Similarly, for a single mode coherent state there is a range of values of the average photon number for which there the generated atom-atom entanglement is AL. In these cases, it was observed that $\ket{eg}$ shows sudden death as squeezing or the average photon numbers were increased.

We have chosen in this analysis to try to isolate the effect of atomic separation on entanglement dynamics from the position-dependent effects arising from boundary conditions.  For this purpose, we have assumed in Sec.~\ref{sec:sysham} an atom-field coupling that depends on the coordinate separating the atoms only by a phase factor.  For clarity we have supposed our two field modes are traveling-wave modes in free space, however a more experimentally relevant situation would be a toroidal resonator, where there is rotational invariance in the azimuthal direction that satisfies our requirements.  In this case the two modes of interest would be two resonant, counter-propagating whispering-gallery modes.  Because strong coupling between an atom and a whispering-gallery modes of a microtoroidal resonator has observed experimentally \cite{KimbleTorus}, there is the possibility of experimentally probing quite directly the model we have considered.  However, in the experimental system there will be dissipative dynamics arising from emission into other modes outside the resonator, so a detailed comparison would require either including the dissipative effects in the theoretical model or a restriction to the case of sufficiently strong coupling to the resonator modes and early times that the dissipation could be neglected.

Our analysis of special cases of the entanglement dynamics arising
in two atoms interacting with two modes suggests a wide variety of
different behaviors can arise, with qualitative features of the
dynamics changing entirely between the two cases, even with the same
initial field state.  This suggests that understanding the distance
dependence of entanglement dynamics for multiple atoms interacting
with a common field will be quite important for predicting even the
qualitative features that may arise. Furthermore, if one has the
practical goal of dynamically generating entanglement or protecting
entanglement once generated, the special cases we have considered
suggest that the ability to achieve these goals will be greatly
impacted by the separation of the atoms.  Having gotten some sense of
the variation in behavior that can arise, the obvious next step would
be to quantify the entanglement dynamics over a significant range of
atomic separations rather than just special or extreme cases. The
variety of qualitative features of entanglement dynamics we have
illustrated here argue for the importance of these considerations.

\acknowledgments
This research is supported in part by grants from the NSA-Laboratory for Physical Sciences, the DARPA-QuEST program (DARPAHR0011-09-1-0008) and  the NSF-ITR program (PHY-0426696).

\end{document}